\documentclass[aps, prl, showpacs, reprint, preprintnumbers,amsmath,amssymb,superscriptaddress,floatfix]{revtex4-2}

\usepackage[]{graphicx}      
\usepackage{bm}             	
\usepackage{times}			
\usepackage{tabularx}
\usepackage{color}
\usepackage{dcolumn}		
\usepackage{amsmath}
\usepackage{amssymb}
\usepackage{amsfonts}
\usepackage{times}
\usepackage{tabularx}
\usepackage{xcolor,colortbl}
\usepackage{amsmath}
\usepackage{tikz,xcolor,hyperref}
\usepackage{blindtext}

\begin{document}
\title{Role of non-reciprocity in spin-wave channeling}
\author{Jean-Paul Adam}
\author{Nathalie Bardou}
\affiliation{Universit\'e Paris-Saclay, CNRS, Centre de Nanosciences et de Nanotechnologies, Palaiseau, France}
\author{Aur\'elie Solignac }
\affiliation{SPEC, CEA, CNRS, Universit\'e Paris-Saclay, 91191 Gif-sur-Yvette, France}
\author{Joo.-Von Kim }
\affiliation{Universit\'e Paris-Saclay, CNRS, Centre de Nanosciences et de Nanotechnologies, Palaiseau, France}

\author{Thibaut Devolder}
\email{thibaut.devolder@cnrs.fr}
\affiliation{Universit\'e Paris-Saclay, CNRS, Centre de Nanosciences et de Nanotechnologies, Palaiseau, France}
       
    \maketitle            
               
\noindent
*: \texttt{thibaut.devolder@cnrs.fr} \\

%
%


%
%

Abstract: The extent to which non-reciprocal waves can be guided in arbitrary directions is an interesting question.
We address one aspect of this problem by studying the propagation of acoustic spin waves in a narrow physical conduit made of a synthetic antiferromagnet. Through a combination of Brillouin Light Scattering microscopy and modeling, we demonstrate that even when attempting to guide waves in the reciprocal direction of the material, the system still exhibits strong signatures of non-reciprocity. This includes the excitation of high wavevector waves in the direction perpendicular to the intended channeling, as well as energy transfer in directions that often neither aligns with the physical conduit nor with the symmetry axes of the magnetic properties. These findings have implications for the modeling of propagating wave spectroscopy in non-reciprocal materials and their potential applications.

\section{Introduction} 
Wave non-reciprocity (NR) -- the alteration of a wave's property upon the reversal of its wavevector -- is a fascinating phenomenon that is seldom observed in linear time-invariant systems \cite{caloz_electromagnetic_2018}, with the notable exception of those involving magnetism. 
The NR of spin-waves (SWs), which can arise from either dipolar interactions \cite{camley_theory_1981} or Dzyaloshinskii-Moriya interaction (DMI) \cite{cortes-ortuno_influence_2013}, has been long regarded as a physical curiosity, primarily enabling a limited number of novel microwave functionalities, or more recently as a handy tool to determine magnetic properties \cite{belmeguenai_interfacial_2015}. However, recent insight has revealed that NR could be significant enough to confer unidirectionality to the energy flow (or to the information flow) carried by SW ensembles \cite{thiancourt_unidirectional_2024}. Moreover, this non-reciprocal behavior can be transferred to other quasiparticles, for instance by appropriate coupling to elastic waves \cite{verba_nonreciprocal_2018, kus_wide-band_2024}. This realization has spurred numerous theoretical and experimental investigations of SW NR in extended films and multilayers \cite{di_enhancement_2015, gladii_frequency_2016, gallardo_reconfigurable_2019, gallardo_spin-wave_2021, matsumoto_large_2022}. 

Theoretical investigations were subsequently conducted in laterally confined geometry \cite{Mruczkiewicz_influence_2016, zingsem_unusual_2019, silvani_effect_2021} primarily focusing on materials exhibiting weak non-reciprocity induced by interfacial DMI. It was predicted that the profiles of the confined SWs would be affected by the NR, meaning that the confined eigenmodes could no longer be represented as a superposition of oppositely propagating SWs with equal wavelengths. Experimental works conducted in stripe geometry \cite{tacchi_suppression_2023} concluded that the frequency asymmetry between counter-propagating spin waves could be even suppressed, suggesting that the confinement may partly restore reciprocity. 

These past studies focused on how NR modifies the eigenmode frequencies and the resulting resonance spectra, yet they overlooked the propagative nature of waves and its potential interest. \textcolor{black}{A first category of studies that fully accounted for the propagative nature of the spin waves utilized Brillouin Light Scattering microscopy in extended media, and analyzed the two-dimensional patterns generated either by diffraction from arrays of slits \cite{makartsou_spin-wave_2024}, or the directional spin wave beams generated by non-uniform transceivers \cite{korner_excitation_2017}. The patterns could be pertinently analyzed with arguments based on iso-frequency theory \cite{lock_properties_2008}}. 
\textcolor{black}{A second category} of studies devoted to SW propagation utilized electrical propagating spin wave spectroscopy (PSWS, \cite{bailleul_propagating_2003}), and involved injecting SWs in wide conduits \cite{thiancourt_unidirectional_2024}, which offered minimal guidance for the waves. The analysis was systematically done using a 1D model \cite{demidov_mode_2008, devolder_propagating-spin-wave_2023} that implicitly assumes that the wavevectors and the group velocity of SWs present in the system are both collinear to the SW conduit.
We will demonstrate that this description is inappropriate when applied to materials with significant NR such a synthetic antiferromagnets (SAFs). Even when the conduit is aligned parallel to the reciprocal direction of a non-reciprocal material, the emission can exhibit features that strongly differ from that encountered when using reciprocal materials. These include for instance the generation of bidirectional nodal structures \textcolor{black}{by a one-dimensional transceiver}, as well as the radiance of a caustics-like energy beam in \textcolor{black}{a \textit{single}} direction that substantially differs from the conduit length.
%
\begin{figure*}
\includegraphics[width=16 cm]{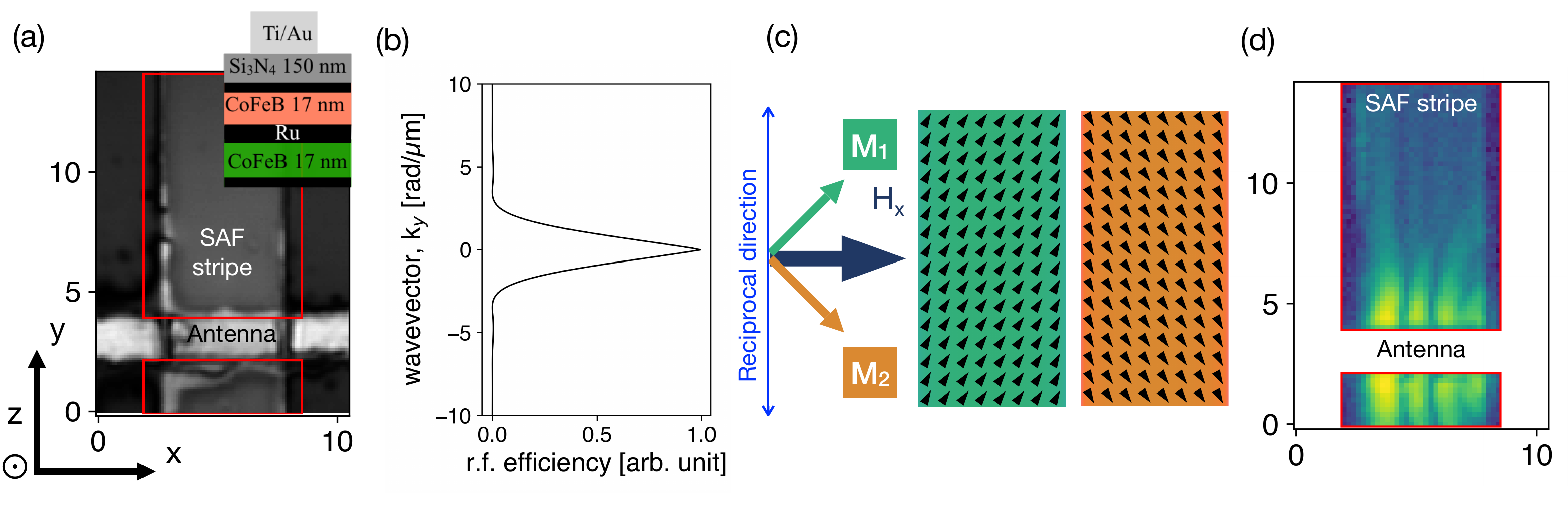}
\caption{\textcolor{black}{Experimental configuration.} (a): Optical image of the device, sketch of the stack and coordinate conventions. \textcolor{black}{The sub-micron vertical gray segments near the left edge of the stripe are resist residues from the fabrication process.} Red boxes: regions investigated by BLS microscopy. (b): Antenna efficiency function versus wavevector. (c): Direction of the applied field, sketch of the resulting scissors state, and micromagnetic configuration at $\mu_0 H_x=50$ mT. Notice the slight non-uniformity of the magnetizations near the edges of the SAF stripe. (d) Example of an experimental BLS microscopy image obtained for $\mu_0 H_x=50$ mT, and a r.f. field oscillating at 4.5 GHz. The image extends 0.8~$\mu$m on both sides of the stripe, whose width is 5~$\mu$m. The top and bottom BLS images have dimensions of 6.6~$\mu$m$\times$10.2~$\mu$m and 6.6~$\mu$m$\times$2.2~$\mu$m, respectively.}
\label{Fig1_conventions_otherLayout}
\end{figure*}
\section{Results}
\subsection{Experimental results} 

We use CoFeB/Ru/CoFeB synthetic antiferromagnetics (SAF), patterned into stripes meant to channel the spin waves. The stripes have widths $w_\textrm{mag} = 5 ~\mu \textrm{m}$ [Fig.~\ref{Fig1_conventions_otherLayout}] and are covered by a microwave antenna of width $w_\textrm{ant}=1.8~\mu \textrm{m}$, able to couple to spin waves of wavevectors satisfying $|k_y| < k_\textrm{y, max}^\textrm{ant}=3~\textrm{rad}/\mu$m. A static field $H_x$ is applied perpendicularly to the stripe. This field orientation induces a situation where the SWs are reciprocal in the direction of the stripe (i.e. $y$) and non-reciprocal in the transverse direction (i.e. $x$). We will analyze the magnetization response when powering the antenna to generate an rf field $\vec{h}^\textrm{rf}_\textrm{ant} \perp \vec H_x$ meant to excite spin waves under the antenna and to channel them in the stripe. This rf field is expected --and was checked \textcolor{black}{(see Fig. 5)}-- to excite the sole acoustic branch of the spin waves of the SAF. 

The magnetization response is imaged by Brillouin Light Scattering microscopy (BLS). The pixel size is $200 \times 200$ nm$^2$ and the optical resolution is estimated to be $2 \pi / k_\textrm{max}^\textrm{BLS} \approx 350$ nm, with $k_\textrm{max}^\textrm{BLS}=18 ~\textrm{rad}/\mu$m being the largest wavevector that can be collected by the BLS set-up (see methods).

%
\begin{figure*}
\centering
\includegraphics[width=\textwidth]{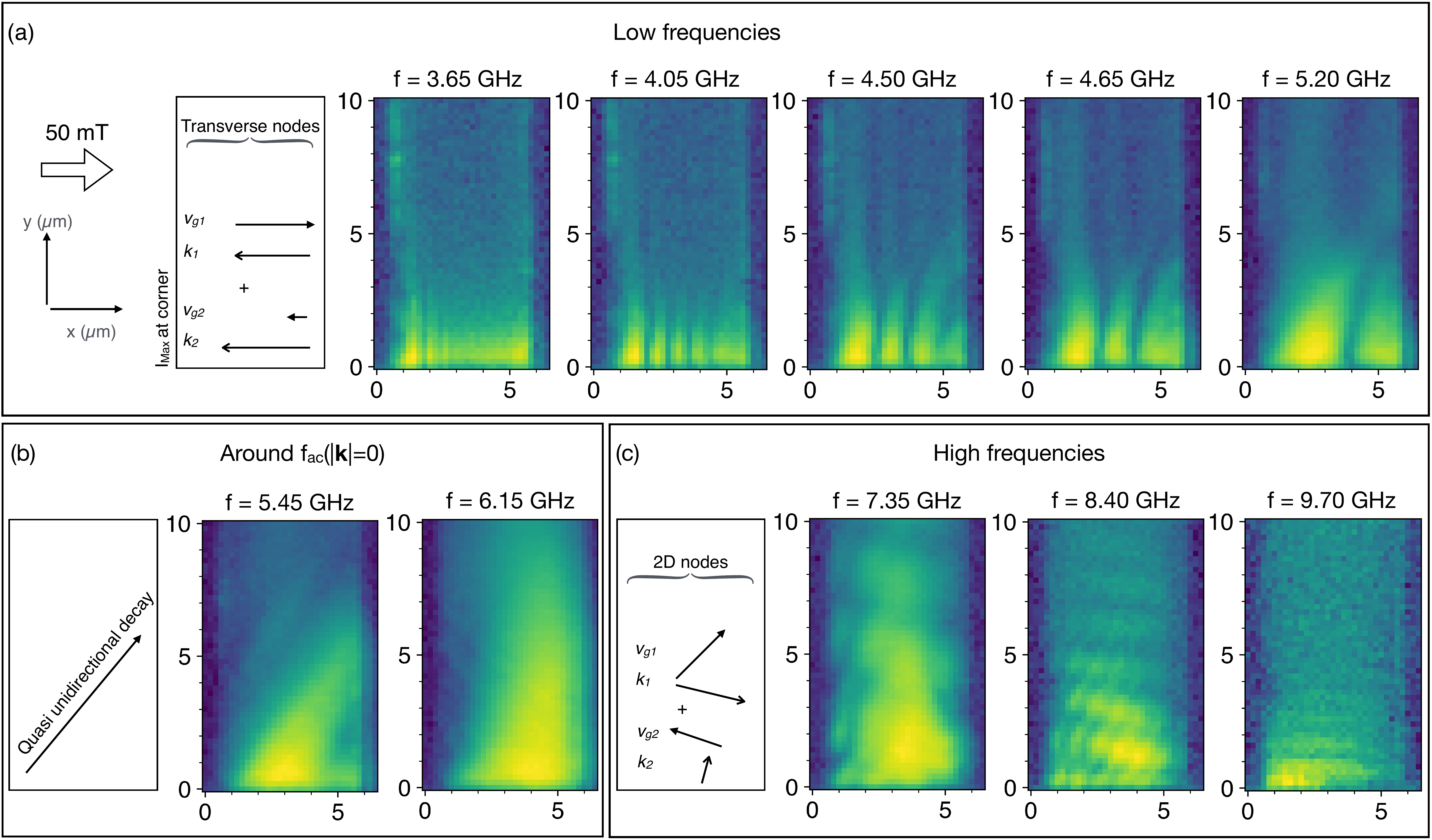}
\caption{Experimental BLS images (logarithmic color scale) taken at an applied field of $\mu_0 H_x$=50~mT, and ordered according to the frequency of the rf signal supplied by the antenna. The images are grouped in 3 types. (a): low frequency patterns with nodes and antinodes in the $x$ direction and intensity maxima at the bottom left corner of the stripe. (b): near ${f_\textrm{ac}(\vec k=\vec 0})$: patterns with tilted energy flow, first approximately in the $\vec x + \vec y$ direction, then progressively rotating towards $\vec y$ when increasing the frequency. (c): high frequency patterns, first exhibiting nodes and antinodes in two directions, then solely in the $\vec y$ direction. The sketches indicates that wavevectors and their group velocities that most contribution to the spatial profile of the BLS signal.}
\label{Figure_2_profiles_of_the_eigenmodes_50mT}
\end{figure*}

Images of the magnetization response were recorded for applied fields of 27, 50, 75 and 100 mT, with qualitatively similar conclusions. Apart from some amplitude non-reciprocity, the magnetization profiles are axially symmetric with respect to the antenna, as in the example of Fig.~\ref{Fig1_conventions_otherLayout}(d); we shall thus now show only the $y>0$ side of the antenna. The profiles of the response can be cast in 3 archetypal categories. The figure ~\ref{Figure_2_profiles_of_the_eigenmodes_50mT} illustrates these categories for the representative 50 mT case, for which the most uniform modes have frequencies in the range $f_\textrm{ac}(\vec k=\vec 0) \approx 5.5-6.5$ GHz . 

At low frequencies (top raw in Fig.~\ref{Figure_2_profiles_of_the_eigenmodes_50mT}), the driven spin waves exhibit nodes and antinodes aligned in the $x$ direction, i.e. transverse to the SAF stripe. The spacing between the nodes gradually increases from 600$\pm 100$ nm at 3.65 GHz to approximately the half stripe with at 5.2 GHz. The BLS signal is confined in the $2~\mu$m vicinity of the antenna: the driven SWs do not seem to radiate energy away from the antenna. Noticeably, the BLS amplitude is maximal near the left edge of the antenna, as if the spin waves were emitted from the bottom left corner of the images towards the bottom right corner. Note that this feature --emission seemly from the left corner-- would never be measured in electrical measurements like propagating spin wave spectroscopy (PSWS), owing to the experimental limitation of only accessing to the spatial average of the magnetization below an antenna. This directionality towards $x>0$ holds strictly at the lowest frequency (3.65 GHz) but some curvature gradually appears when increasing the frequency (see e.g. the 4.65 and 5.2 GHz patterns in Fig.~\ref{Figure_2_profiles_of_the_eigenmodes_50mT}). 

Approaching the frequency of the uniform acoustic SWs disrupts the pattern of dynamic magnetization 
[5.45 GHz and subsequent frames in Fig.~\ref{Figure_2_profiles_of_the_eigenmodes_50mT}(b)]:
the antinodes are expelled from the stripe, and the pattern rotates as a whole: it seems that the SWs are now emitted by the antenna in a tilted directional manner. The direction of decay of the BLS intensity has become approximately $\vec x + \vec y$. Because of this $\approx$45 degree directional emission, the BLS signal vanishes at a distance of $w_\textrm{ant}$ away from the antenna, exactly when the caustics-like SW beam hits the right edge of the stripe. Energy is indeed radiated away from the antenna but the BLS signal does not decay exponentially along the stripe length. Noticeably, this indicates that the usual way \cite{pirro_spin-wave_2014, demidov_excitation_2016} of deducing SW attenuation length becomes inadequate in the joint presence of non-reciprocity and confinement. 
A slight further increase of the frequency [6.15 GHz case, Fig.~\ref{Figure_2_profiles_of_the_eigenmodes_50mT}(b)] rotates the decay direction to the (more conventional) $\vec y$ direction, with a long propagation distance comparable to the image size (10.2 $\mu$m). In this narrow frequency range the decay direction is the one expected from a channeling by the stripe; however the material non-reciprocity still plays a role and breaks the $\{x, -x\}$ symmetry: the maximum intensity is not at the midst of the SAF stripe but rather offset by 1 $\mu$m to the right.

In the 7-9 GHz range [Fig.~\ref{Figure_2_profiles_of_the_eigenmodes_50mT}(c)], a bidimensional array of nodes and antinodes reappears progressively, with beatings in the two  $x$ and $y$ directions, indicating the coexistence of several modes with unequal wavevectors. The overall decay of the BLS signal stays essentially in the $y$ direction, still with a long attenuation length. The density of nodes and antinodes increases with the frequency. Finally at the highest investigated frequencies (9.7 GHz), we recover a behavior resembling the low frequency one, i.e. with an apparent emission from the bottom left corner of the SAF stripe, but with some beating along the $y$ direction.

\subsection{Physical understanding}
Several features of these BLS profiles are quite unusual. They include:\\ (i) the systematic breaking of the $\{x, -x\}$ symmetry that witnesses the material non-reciprocity 
and, \\
(ii) the apparent ability of an antenna that generates a field with translational invariance in the $x$ direction to excite magnetization profiles with a surprisingly large number of nodes in this same direction. 


In past studies of the SWs in confined geometries, it has often been possible to understand the nature of the modes by using the dispersion relation in the corresponding unbounded films and applying some quantification of the transverse wavevector \cite{bayer_spin-wave_2006} according to specific boundary conditions \cite{guslienko_effective_2002}. The BLS images could then be understood from overlap integrals of the antenna r.f. field and the profiles of the few allowed modes \cite{demidov_magnonic_2015}. 
\subsection{Dispersion relations}

Let us apply the same rational for the SAF stripes with the dispersion relations reported in Fig.~\ref{Figure_3_AcousticDispersionRelations_v3} and calculated in the dynamical matrix formalism (see methods). The dispersion relation in Fig.~\ref{Figure_3_AcousticDispersionRelations_v3} recalls that as well-known for SAFs in the scissors state, the acoustic spin waves are reciprocal in the $k_y$ direction [inducing $\omega_\textrm{ac}(k_x, k_y)=\omega_\textrm{ac}(k_x, -k_y)$] while being non-reciprocal and unidirectional in the field direction (i.e. for small $k_x$, we have $\frac{\partial\, \omega_\textrm{ac}} {\partial\, k_x} \Big \rvert_{k_y=0} > 0 $). 
For the applied fields considered here, this unidirectional character extends to far above the region $[-k_{y, \textrm{ant}}^\textrm{max}, -k_{y,\textrm{ant}}^\textrm{max}]$ that can be addressed by the antenna, except for modes with very large $-k_{x}$ that can anyway not be detected by BLS (Fig.~\ref{Figure_3_AcousticDispersionRelations_v3}). 
\subsection{Interpretation guidelines} 

The presence of several to many nodes in the $x$ direction argues for the generation of SWs of wavevectors with arbitrarily large $k_x$ components. This requires non-uniformity in the $x$ direction at scales compatible with all these $k_x$'s, hence at lateral scales smaller than $2 \pi / k_\textrm{max}^\textrm{BLS}$. These non-uniformities are likely the tiny regions near the stripe edges where the equilibrium magnetization and its demagnetizing fields are non-uniform [see Fig.~\ref{Fig1_conventions_otherLayout}(c)]. 

\begin{figure*}
\includegraphics[width=8 cm]{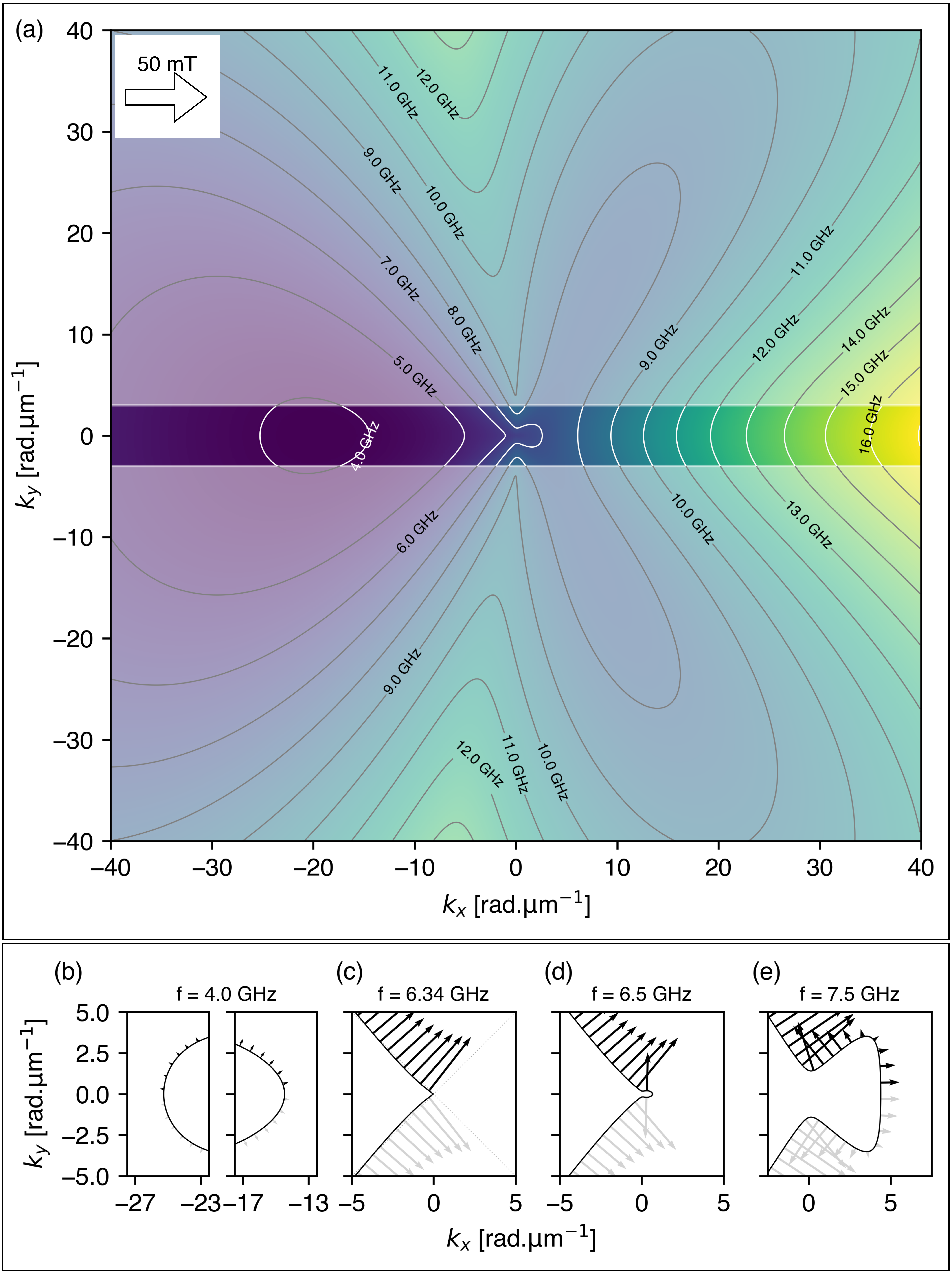}
\caption{Dispersion relations of the acoustic spin waves in an infinitely extended SAF film at a field of $\mu_0 H_x$=50~mT. (a): Dispersion according to the dynamical matrix formalism. The shaded areas are the regions in which the antenna cannot detect spin waves. Sketched dispersion relations at low frequency (b), at $\vec k= \vec 0$ frequency (c), slightly above (d) and much above (e). The  arrows sketch the directions of the group velocities. The arrows are grey when the group velocity points downward from the antenna.}
\label{Figure_3_AcousticDispersionRelations_v3}
\end{figure*}

Besides, we can interpret the BLS images by considering only the wavevectors whose group velocity is directed to the upper side (i.e. $y>0$) of the antenna, since the other SWs cannot radiate energy into the region under evaluation. 
Since the driven SWs must have a $k_y$ component that can couple to the antenna spectrum [Fig.~\ref{Fig1_conventions_otherLayout}(b)], the relevant wavevector space to be considered for the qualitative analysis of the BLS images is thus: 
\begin{equation}
\left\{ \begin{array}{l} 
\{k_x, k_y\} \in [-k_{\textrm{BLS}}^\textrm{max}, k_{\textrm{BLS}}^\textrm{max}]\times [-k_{y,\textrm{ant}}^\textrm{max}, k_{y, \textrm{ant}}^\textrm{max}]  \\
\textrm{~~~and~~~~} (\vec{\nabla}_{\vec k} \omega) . \vec{y} > 0 \\
\end{array} \right.\label{truncation}
\end{equation}
For each applied frequency $\omega_\textrm{applied} / (2 \pi)$, we shall thus analyze the BLS images by considering the corresponding isofrequency contour $\omega_\textrm{ac}(k_x, k_y)=\omega_\textrm{applied}$ after truncation of the spin wave manifold according to Eq.~\ref{truncation}. 
The complicated shape of the isofrequency contours (Fig.~\ref{Figure_3_AcousticDispersionRelations_v3}) provides the second interpretation guideline: the large differences in spatial patterns among the BLS images 
are likely to result from the mutual orientation of the wavevector, the group velocities ($\vec v_g = \vec{\nabla}_{\vec k} \omega)$, the antenna, and  the stripe-shaped SW conduit.

%
\subsection{Qualitative understanding of the experimental BLS images from the dispersion relation} 

At low frequencies (4 GHz), the relevant SWs almost share the same group velocity pointing in the $\vec x$ direction and the same large ($k_x^{4\,\textrm{GHz}} \approx -14~\textrm{rad}/\mu$m) value of $k_x$. They only differ by their --comparatively much smaller-- $k_y$ value. 
Because of their common $v_g$ direction, the amplitude of the signal of every SW present in the system decays in the $\vec x$ direction. This explains why the bottom left corner of the BLS images at low frequency appeared systematically as the brightest spot in experiments (see the top left images in Fig.~\ref{Figure_2_profiles_of_the_eigenmodes_50mT}). Note that the SWs are emitted \textit{in-phase} at every position \textcolor{black}{$x_0 \in [-\frac {w_\textrm{stripe}}{2}, \frac {w_\textrm{stripe}}{2}]$} along the antenna, but each of these SWs has a \textcolor{black}{spatially-dependent} phase factor \textcolor{black}{$e^{i k_x^{4\,\textrm{GHz}} (x-x_0)}$ and an amplitude decay factor $ \propto e^{- ((x+w/2)/L_\textrm{att})} \mathcal{H}(x-x_0)$, where $\mathcal{H}$ is the heaviside function and $L_\textrm{att}$ the attenuation length}. The phases of these spin waves rotate along the $\vec x$ direction at the common spatial pace $k_x^{4\,\textrm{GHz}}$. \textcolor{black}{The interference of all these waves leads to a BLS pattern whose envelope is maximal at the left edge and decays like $L_\textrm{att}$. Interference of the profiles of these waves is constructive whenever they are emitted at $x_0$ positions spaced by integers of $\delta x = 2 \pi /|k_x^{4\,\textrm{GHz}}| \approx 450$ nm away from the left edge. The destructive interference will lead to nodes in the BLS images, and antinodes otherwise, in reasonable agreement with the experimental node-to-node spatial periodicity.} 

At 6.34 GHz (which is the uniform resonance of the acoustic mode in an extended film), the truncated isofrequency contour is line-shaped and it runs along approximately the $k_x=- k_y$ diagonal direction, as sketched in Fig.~\ref{Figure_3_AcousticDispersionRelations_v3}(c). All the SWs that can be excited at this frequency share the same direction of group velocity, pointing approximately in the $\vec x +\vec y$. These SWs span over a continuous range of wavevectors from $\vec 0$ to $k_\textrm{max}^\textrm{ant}(\vec y - \vec x)$. The broadband character of the wavevector spectrum is such that no fine structure is expected in the spatial domain. However, the common group velocity ensures that every generated SW decays in the same $\vec x +\vec y$ diagonal direction. This is the reason why a very well defined $\approx$45$^\circ$ directional pattern was observed experimentally in this frequency range [Fig.~\ref{Figure_2_profiles_of_the_eigenmodes_50mT}(b)]. 

When slightly increasing the applied frequency to 6.5 GHz, a semi-circular protuberance starts to stick out of the angled part of the isofrequency contour [Fig.~\ref{Figure_3_AcousticDispersionRelations_v3}(d)]. The SWs whose wavevector belongs to this protuberance have group velocities that lie out of the approximate $\vec x +\vec y$ direction and span from $+\vec x$ to $+\vec y$. This growth of the protuberance is expected to weaken the directional character of the BLS image that was originating from the line-shaped $\approx 45^\circ$-oriented segment of the isofrequency contour. In experiments, the directionality was indeed progressively lost as increasing the frequency slightly above the frequency of the uniform acoustic spin wave resonance. 

At 7.5 GHz [Fig.~\ref{Figure_3_AcousticDispersionRelations_v3}(e)], the protuberance takes all the relevant wavevector space while distorting to get a trapezoidal shape. The driven SWs have now wavevectors that are broadly distributed, but with two well-defined subgroups of group velocities, heading either along approximately $\vec x +\vec y$ like before, or along approximately $\vec y -\vec x$. This two-flavor distribution is expected to lead to beating patterns in these two directions 
 as was indeed anticipated from the experimental result. [Fig.~\ref{Figure_2_profiles_of_the_eigenmodes_50mT}(d)].
Note that if we were performing electrical propagating spin wave spectroscopy in such a narrow spin wave conduit, the conventional analysis of PSWS data \cite{talmelli_reconfigurable_2020} would fail because it requires a 1D flow of a single family of SWs, a condition that is obviously not satisfied here.

Finally at high frequency (10 GHz for instance), we recover a situation that bears some similarity with the low frequency case: the isofrequency contour is almost a vertical line, such that the relevant SWs share a large common wavevector $k_x^{10\,\textrm{GHz}} \approx +13~\textrm{rad}/\mu$m and a common group velocity pointing towards $x$. With the same rational as for the low frequency case, this results in a line of nodes and antinodes of decaying amplitude from the bottom left corner to the bottom right corner of the BLS images. The node-to-node distance is expected to be now $2 \pi / k_x^{10\,\textrm{GHz}} =480$ nm, close to the resolution of our images, hence difficult to detect experimentally [Fig.~\ref{Figure_2_profiles_of_the_eigenmodes_50mT}(e)].

\section{Discussion}
In partial conclusion, qualitative understanding of the BLS images is possible from the thin-film dispersion relation, at least in situations selected for their simplicity. However it cannot account for the details of the images, notably when the isofrequency contours are very curvy and the node/antinode pattern runs in two different directions. Also, the above analysis relies on plane waves, such that it does not account for demagnetizing effects leading to a difference between an unbounded SAF and a patterned stripe, where the internal field is non-uniform and lower than the applied field. Finally, it misses any possible hybridation between the optical and the acoustic spin waves. It has been indeed shown that there are regions of the parameter space where the two SW branches cross each other [i.e. when $\omega_\textrm{ac}(k_x, k_y, H_x) \approx\omega_\textrm{op}(k_x, k_y, H_x)$] thereby hybridizing the optical and acoustic characters \cite{shiota_tunable_2020}. As a result, SWs with a partial optical character may in principle be weakly driven by the antenna in this region of the parameter space. These deficiencies of the qualitative rational can be resolved by calculating the response from micromagnetics for a quantitative comparison with the experimental results.

%
\begin{figure*}
\includegraphics[width=\textwidth]{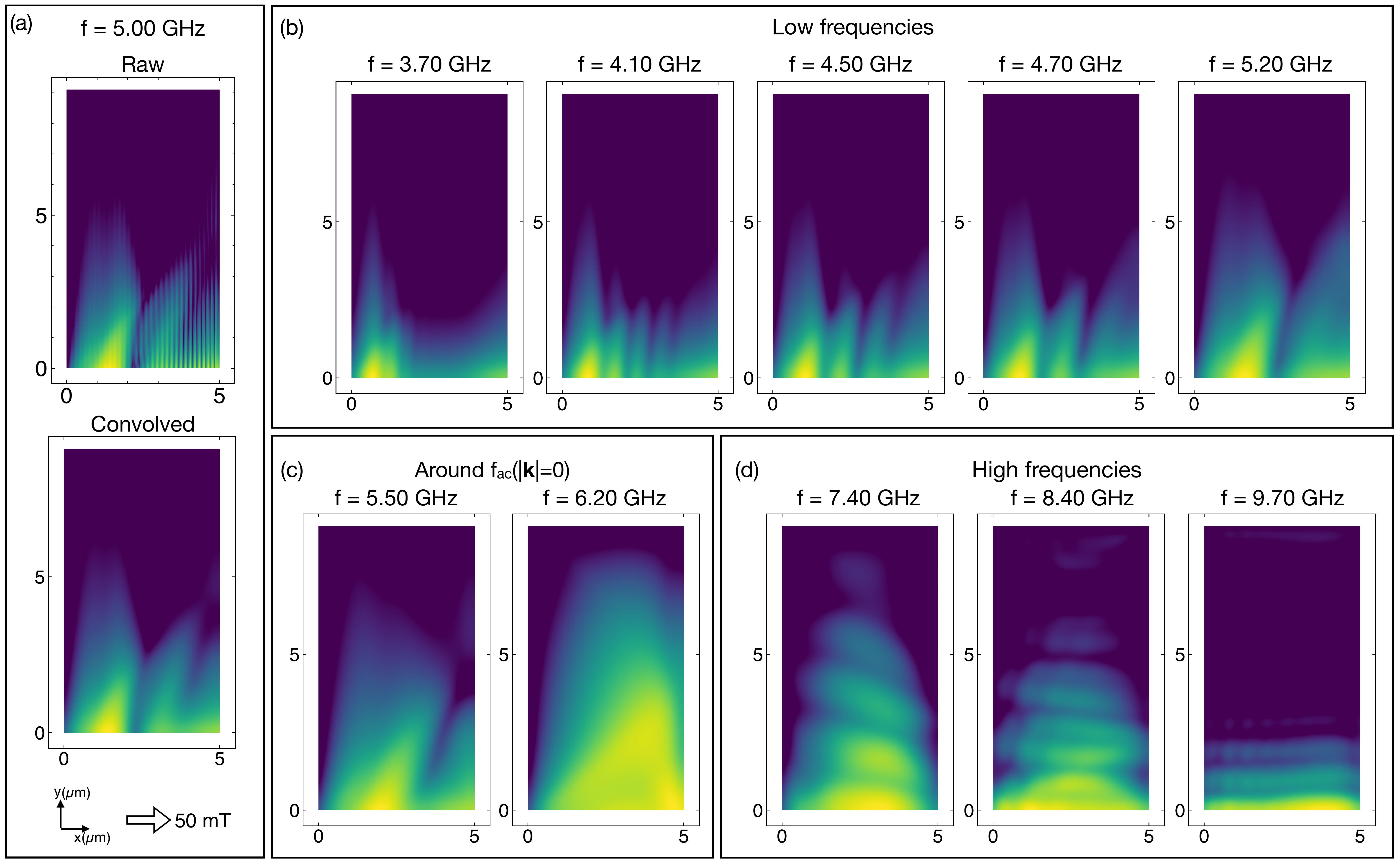}
\caption{Micromagnetic simulations of the BLS images for an infinite lateral resolution (Eq.~\ref{HowToCalculateBLSimages}, logarithmic color scale) and after convolution with a Gaussian filter reducing the lateral resolution to 350 nm. The bottom of the images ($y=0$) are aligned with the outer edge of the antenna. The calculation are done for a field of $\mu_0$H$_x$=50~mT and mirror the experimental behaviors reported in Fig.~\ref{Figure_2_profiles_of_the_eigenmodes_50mT}. (a): Low frequency patterns with nodes and antinodes that alternate in the $x$ direction. (b), near $\omega_\textrm{ac}(\vec k=\vec 0)$: patterns with a tilted flow of energy in the $\vec x + \vec y$ direction. (c): patterns measured in the intermediate frequencies, exhibiting nodes and antinodes in two directions.}
\label{Figure_4_micromagneticResults}
\end{figure*}


The Fig.~\ref{Figure_4_micromagneticResults} reports a selection of BLS images predicted by micromagnetics (see Eq.~\ref{HowToCalculateBLSimages} in methods) \textcolor{black}{for our stripe width of 5 $\mu$m. Calculations of other stripe widths from 1 to 10 $\mu$m are reported in \textcolor{black}{Fig. 7}}. If convolved with the optical resolution of the BLS, the agreement with their experimental counterparts (Fig.~\ref{Figure_2_profiles_of_the_eigenmodes_50mT}) is remarkable, but there are several small differences that deserve a discussion. 

In the low frequency case [see Fig.~\ref{Figure_4_micromagneticResults}(a) \textcolor{black}{as well as the data for other stripe widths in the \textcolor{black}{Fig. 7 and 8}}], the micromagnetic calculations indicate that, in addition to the line of nodes and antinodes seemingly emitted from the bottom left corner already identified in experiments, another line of nodes and antinodes seems emitted from the bottom right corner. This second line of nodes has a substantially smaller spatial period and could not be resolved in the experiments. However when convolved with the optical resolution of the BLS, this second line of nodes results in a secondary maximum of BLS intensity at the bottom right corner, which was indeed perceived experimentally [Fig.~\ref{Figure_2_profiles_of_the_eigenmodes_50mT}(a)]. \\This second line of nodes can be understood with the same rational as the first one, by noticing that the isofrequency contours at low frequency have egg shapes and thus cross Eq.~\ref{truncation} twice [Fig.~\ref{Figure_3_AcousticDispersionRelations_v3}(a)]. For instance the 4 GHz contour extends far in the $k_x<0$ direction and passes at the point $\{k_x^{4\textrm{GHz'}}, k_y\}=\{-25, 0\}$ rad/$\mu$m., where the group velocity points in the $-\vec x$ direction. This wavevector and its group velocity is responsible for the line of antinodes of periodicity $2 \pi / k_x^{4\textrm{GHz'}}$, decaying leftwards from the bottom right corner of the BLS images. 

This pattern of nodes and antinodes progressively expands when increasing the applied frequency. When approaching the frequency of the uniform mode [Fig.~\ref{Figure_4_micromagneticResults}(b)] there remains a clear directional pattern with a decay approximately in the $\vec x + \vec y$ direction, which reproduces nicely the experiment and the expectations drawn from the dispersion relation. \textcolor{black}{This directional pattern is also predicted for other stripe widths, see \textcolor{black}{Fig. 7}.}
However at 8.4 GHz [Fig.~\ref{Figure_4_micromagneticResults}(c)], the micromagnetic simulation fails to reproduce the bidirectional beating pattern observed experimentally and that could be understood from the shape of the protuberance within the dispersion relation. The absence of the modulation in the $x$ direction in the micromagnetics result at 8.4 GHz is probably indicative that the inhomogeneity of the magnetization distribution in the $x$ direction (near the stripe edges) is not correctly modeled and does not allow to populate the second flavor of modes that appears at after the blossoming of the protuberance. 

\textcolor{black}{Finally, the evolution of the predicted BLS images for very narrow stripes is worth discussing.  The non-reciprocity of the spin waves in the relevant wavevector space (Eq.~1) is progressively lost when the stripe width gets smaller than the antenna width. For a 1 micron stripe width (see \textcolor{black}{Fig.~8)}, we recover a situation close to the well-known case of a single-layer magnet in the (reciprocal) Damon-Eshbach case, where interference occurs between the uniform mode and the higher order confined spin waves, leading to periodic self-modulation of the amplitude pattern \cite{demidov_mode_2008}. }



\textbf{
}

In summary, we have used Brillouin Light Scattering microscopy and micrometer-sized rf antennas to study the confinement and the propagation of the acoustic spin waves in narrow conduits made of synthetic antiferromagnets that exhibit a giant non-reciprocity (NR). One could think that when the SWs are injected in a conduit that is parallel to the reciprocal direction, the NR would not play a major role. The contrary happens: the non-reciprocal nature of the dispersion relation is such that the direction in which a line-shaped antenna radiates SW energy rarely matches with the conduit direction. The most striking consequence of NR is that when below the frequency of uniform acoustic SWs, the SWs driven by the antenna have wavevectors running parallel to the antenna, i.e. perpendicular to the expected guiding direction.
This leads to a unique nodal structure of the SW pattern in the vicinity of the antenna, with no possible analogy to reciprocal situations. 

Electrical spectroscopy of propagating spin waves has recently become a popular tool for the study of spin waves. It involves measuring the magnetization response, but spatially averaged below a receiving antenna. Since imaging a system's response is key to understand such a spatially averaged response, our results are essential to interpret PSWS experiments done on non-reciprocal systems and avoid errors. 
Beyond magnonics, our results can be used to anticipate which modes can be harnessed for the transport of energy/information in conduits when using non-reciprocal waves.

%

\section*{Methods}
\subsection{Devices} 
We use in-plane magnetized Synthetic AntiFerromagnetic (SAF) films of composition Co$_{40}$Fe$_{40}$B$_{20}$ (17~nm) /Ru (0.7~nm) / Co$_{40}$Fe$_{40}$B$_{20}$ (17~nm) [Fig.~\ref{Fig1_conventions_otherLayout}(a)]. 
The SAF film was patterned into stripes of width $w_\textrm{mag} = 5 ~\mu \textrm{m}$ [Fig.~\ref{Fig1_conventions_otherLayout}(b)]. We then deposited a $s=150$ nm-thick insulation layer of Si$_3$N$_4$ and a single-wire antenna of width $w_\textrm{ant}=1.8~\mu \textrm{m}$ and thickness $h=160$ nm. A static field $H_x$ is applied in the direction transverse to the stripe. The films, devices and variants thereof were characterized extensively by ferromagnetic resonance \cite{mouhoub_exchange_2023} and propagating spin wave spectroscopy \cite{thiancourt_unidirectional_2024} to deduce the material properties. This include magnetization $M_s=1.35$ MA/m, damping $\alpha=0.011$, exchange stiffness $A$ = 16 pJ/m, and interlayer coupling $J=-1.0$ mJ/m$^2$. \textcolor{black}{These values were concluded from the analysis of the 4 lowest frequency spin wave modes of an unpatterned SAF, as measured by VNA-FMR \cite{mouhoub_exchange_2023}.}

The experiments shall analyze the magnetization response when feeding the antenna with an rf source. The field of the antenna obeys $\vec{h}^\textrm{rf}_\textrm{ant} \perp \vec H_x$ such that it should excite the acoustic spin waves. However it would not excite the optical spin waves if the SAF was an infinitely extended film. We show in the \textcolor{black}{Fig.~5} that this also holds in our stripe despite of the lateral confinement. Besides, at the applied fields used in this study, the optical spin waves have frequencies above our investigated frequency interval, as shown in \textcolor{black}{Fig.~6}. We will thus primarily analyse our experiments by considering the sole acoustic branch of the spin wave manifold.

\subsection{Brillouin Light Scattering measurements} 
Our goal is to understand how the lateral confinement within the SAF stripe affects the SWs and their propagation, for fields applied in the width direction of the stripe. 
The spectra of the magnetization response and their spatial profiles are measured by Brillouin Light Scattering (BLS).  
The images in Fig.~\ref{Fig1_conventions_otherLayout}(d), and \ref{Figure_2_profiles_of_the_eigenmodes_50mT} are taken in microscopy imaging configuration and in driven mode: the laser spot is scanned across the sample, and the antenna is fed by a monochromatic continuous r.f. source. The pixel size is $200 \times 200$ nm$^2$ and the optical resolution is estimated to be $2 \pi / k_\textrm{max}^\textrm{BLS} \approx 350$ nm, with $k_\textrm{max}^\textrm{BLS}=18 ~\textrm{rad}/\mu$m being the largest wavevector that can be collected by the experimental set-up. 

\subsection{Micromagnetic simulations} 
Our micromagnetic simulations are done using the mumax3 code at zero temperature \cite{vansteenkiste_design_2014}. For each applied field $H_x$, we start by energy minimization to converge to the ground state, which is a scissors state. The magnetizations are uniform in the central part of the stripe but tend to be more parallel to the edges of the stripe in the regions near the edges of the stripe [see Fig.~\ref{Fig1_conventions_otherLayout}(c)].

The theoretical dynamical response of the stripe is evaluated by computing the steady state response of the magnetization to the two components $h_y^\textrm{rf}(y, \forall x)$ 
and $h_z^\textrm{rf}(y, \forall x)$ of the antenna field.  
The spatial profile of the r.f. field is calculated for the real geometry (Eqs.~A1-A4 of ref.~\onlinecite{devolder_propagating-spin-wave_2023}) and a peak field of $h_y^\textrm{rf}=1$~mT at the SAF surface under the middle of the antenna. The power spectral density of $h_y^\textrm{rf}(k_y)$ in wavevector space is displayed in Fig.~\ref{Fig1_conventions_otherLayout}(b). It has a substantial amplitude up to $|k_y| < k_\textrm{y, max}^\textrm{ant}=3~\textrm{rad}/\mu$m. 

We assume that the BLS signal comes essentially \cite{hamrle_analytical_2010} from the square of the out-of-plane (i.e $m_z$) component of the magnetization in this steady state: we thus construct the expected BLS microscopy images by the following variance: 
\begin{equation}
    \textrm{BLS}(x, y)=\int_\textrm{period} \Big(m_z(x, y, t)-m_z(x, y, t=0)\Big)^2 dt \label{HowToCalculateBLSimages}
\end{equation}
To account for the finite lateral resolution of BLS microscopy, a Gaussian filter will be applied to expected images.
\subsection{Theory} 
The images will be discussed from a thorough examination of the dispersion relations 
of the spin waves of an unbounded SAF, which we obtained using the formalism of the dynamical matrix \cite{Nortemann_microscopic_1993, Grimsditch_magnetic_2004} \textcolor{black}{with explicit accounting of the interlayer dipole-dipole interactions using Eqs. 26 and 32 of ref.~\onlinecite{henry_propagating_2016}. We describe each of the two layers of the SAF by discretizing it into two slices. The key ingredient for the non-reciprocity is the dipole-dipole interaction between the dynamic magnetizations of each layer of the SAF: the dynamic magnetization associated with a spin wave rotates in space, creating a stray field that is stronger at one side of the layer, this side being defined by the sign of the wavevector (see Fig.~4 of ref.~\onlinecite{devolder_propagating-spin-wave_2023}). This effectively couples more the two layers for one sign of the wavevector, creating the non-reciprocity. The non-reciprocity scales with the thickness and the cosine of the wavevector direction (Eq.~3 of ref.~\onlinecite{millo_unidirectionality_2023})}. 
For each applied field, the calculations are run for a grid of wavevectors $\{k_x, k_y\}$ and up to $\pm 40 ~\textrm{rad}/\mu$m.
The two lowest order families of SAF spin waves are calculated: the acoustic (ac) SWs of frequencies $\omega_\textrm{ac}$, and the optical (op) SWs. A representative example of the dispersion relation of the acoustic SWs is reported in Fig.~\ref{Figure_3_AcousticDispersionRelations_v3}.

As well-known for SAFs in the scissors state \textcolor{black}{(see the part on acoustic SW modes in Table I, in \cite{millo_unidirectionality_2023})}, the acoustic spin waves are reciprocal in the $k_y$ direction [inducing $\omega_\textrm{ac}(k_x, k_y)=\omega_\textrm{ac}(k_x, -k_y)$] while being non-reciprocal and unidirectional in the field direction (i.e. $\forall k_x, \frac{\partial\, \omega_\textrm{ac}} {\partial\, k_x} \Big \rvert_{k_y=0} > 0 $). For the applied fields considered here, this unidirectional character extends to far above the region $[-k_{y, \textrm{ant}}^\textrm{max}, -k_{y,\textrm{ant}}^\textrm{max}]$ that can be addressed by the antenna, except for modes with very large $-k_{x}$ that can anyway not be detected by BLS (Fig.~\ref{Figure_3_AcousticDispersionRelations_v3}).


\section*{ACKNOWLEDGEMENTS}
 This work was supported by public grants overseen by the French National Research Agency (ANR) as part of the “Investissements d'Avenir” and France 2030 programs (Labex NanoSaclay, reference: ANR-10-LABX-0035, project SPICY. Also: PEPR SPIN, references: ANR 22 EXSP 0008 and ANR 22 EXSP 0004). T.D. acknowledges the French National Research Agency (ANR) under Contract No. ANR-20-CE24-0025 (MAXSAW). This work was supported by the RENATECH network. We thank S. M. Ngom for sample fabrication.




\begin{thebibliography}{36}%
\makeatletter
\providecommand \@ifxundefined [1]{%
 \@ifx{#1\undefined}
}%
\providecommand \@ifnum [1]{%
 \ifnum #1\expandafter \@firstoftwo
 \else \expandafter \@secondoftwo
 \fi
}%
\providecommand \@ifx [1]{%
 \ifx #1\expandafter \@firstoftwo
 \else \expandafter \@secondoftwo
 \fi
}%
\providecommand \natexlab [1]{#1}%
\providecommand \enquote  [1]{``#1''}%
\providecommand \bibnamefont  [1]{#1}%
\providecommand \bibfnamefont [1]{#1}%
\providecommand \citenamefont [1]{#1}%
\providecommand \href@noop [0]{\@secondoftwo}%
\providecommand \href [0]{\begingroup \@sanitize@url \@href}%
\providecommand \@href[1]{\@@startlink{#1}\@@href}%
\providecommand \@@href[1]{\endgroup#1\@@endlink}%
\providecommand \@sanitize@url [0]{\catcode `\\12\catcode `\$12\catcode
  `\&12\catcode `\#12\catcode `\^12\catcode `\_12\catcode `\%12\relax}%
\providecommand \@@startlink[1]{}%
\providecommand \@@endlink[0]{}%
\providecommand \url  [0]{\begingroup\@sanitize@url \@url }%
\providecommand \@url [1]{\endgroup\@href {#1}{\urlprefix }}%
\providecommand \urlprefix  [0]{URL }%
\providecommand \Eprint [0]{\href }%
\providecommand \doibase [0]{https://doi.org/}%
\providecommand \selectlanguage [0]{\@gobble}%
\providecommand \bibinfo  [0]{\@secondoftwo}%
\providecommand \bibfield  [0]{\@secondoftwo}%
\providecommand \translation [1]{[#1]}%
\providecommand \BibitemOpen [0]{}%
\providecommand \bibitemStop [0]{}%
\providecommand \bibitemNoStop [0]{.\EOS\space}%
\providecommand \EOS [0]{\spacefactor3000\relax}%
\providecommand \BibitemShut  [1]{\csname bibitem#1\endcsname}%
\let\auto@bib@innerbib\@empty
\bibitem [{\citenamefont {Caloz}\ \emph {et~al.}(2018)\citenamefont {Caloz},
  \citenamefont {Alù}, \citenamefont {Tretyakov}, \citenamefont {Sounas},
  \citenamefont {Achouri},\ and\ \citenamefont
  {Deck-Léger}}]{caloz_electromagnetic_2018}%
  \BibitemOpen
  \bibfield  {author} {\bibinfo {author} {\bibfnamefont {C.}~\bibnamefont
  {Caloz}}, \bibinfo {author} {\bibfnamefont {A.}~\bibnamefont {Alù}},
  \bibinfo {author} {\bibfnamefont {S.}~\bibnamefont {Tretyakov}}, \bibinfo
  {author} {\bibfnamefont {D.}~\bibnamefont {Sounas}}, \bibinfo {author}
  {\bibfnamefont {K.}~\bibnamefont {Achouri}},\ and\ \bibinfo {author}
  {\bibfnamefont {Z.-L.}\ \bibnamefont {Deck-Léger}},\ }\bibfield  {title}
  {\bibinfo {title} {Electromagnetic {Nonreciprocity}},\ }\href
  {https://doi.org/10.1103/PhysRevApplied.10.047001} {\bibfield  {journal}
  {\bibinfo  {journal} {Physical Review Applied}\ }\textbf {\bibinfo {volume}
  {10}},\ \bibinfo {pages} {047001} (\bibinfo {year} {2018})}\BibitemShut
  {NoStop}%
\bibitem [{\citenamefont {Camley}\ \emph {et~al.}(1981)\citenamefont {Camley},
  \citenamefont {Rahman},\ and\ \citenamefont {Mills}}]{camley_theory_1981}%
  \BibitemOpen
  \bibfield  {author} {\bibinfo {author} {\bibfnamefont {R.~E.}\ \bibnamefont
  {Camley}}, \bibinfo {author} {\bibfnamefont {T.~S.}\ \bibnamefont {Rahman}},\
  and\ \bibinfo {author} {\bibfnamefont {D.~L.}\ \bibnamefont {Mills}},\
  }\bibfield  {title} {\bibinfo {title} {Theory of light scattering by the
  spin-wave excitations of thin ferromagnetic films},\ }\href
  {https://doi.org/10.1103/PhysRevB.23.1226} {\bibfield  {journal} {\bibinfo
  {journal} {Physical Review B}\ }\textbf {\bibinfo {volume} {23}},\ \bibinfo
  {pages} {1226} (\bibinfo {year} {1981})}\BibitemShut {NoStop}%
\bibitem [{\citenamefont {Cortés-Ortuño}\ and\ \citenamefont
  {Landeros}(2013)}]{cortes-ortuno_influence_2013}%
  \BibitemOpen
  \bibfield  {author} {\bibinfo {author} {\bibfnamefont {D.}~\bibnamefont
  {Cortés-Ortuño}}\ and\ \bibinfo {author} {\bibfnamefont {P.}~\bibnamefont
  {Landeros}},\ }\bibfield  {title} {\bibinfo {title} {Influence of the
  {Dzyaloshinskii}–{Moriya} interaction on the spin-wave spectra of thin
  films},\ }\href {https://doi.org/10.1088/0953-8984/25/15/156001} {\bibfield
  {journal} {\bibinfo  {journal} {Journal of Physics: Condensed Matter}\
  }\textbf {\bibinfo {volume} {25}},\ \bibinfo {pages} {156001} (\bibinfo
  {year} {2013})}\BibitemShut {NoStop}%
\bibitem [{\citenamefont {Belmeguenai}\ \emph {et~al.}(2015)\citenamefont
  {Belmeguenai}, \citenamefont {Adam}, \citenamefont {Roussigné},
  \citenamefont {Eimer}, \citenamefont {Devolder}, \citenamefont {Kim},
  \citenamefont {Cherif}, \citenamefont {Stashkevich},\ and\ \citenamefont
  {Thiaville}}]{belmeguenai_interfacial_2015}%
  \BibitemOpen
  \bibfield  {author} {\bibinfo {author} {\bibfnamefont {M.}~\bibnamefont
  {Belmeguenai}}, \bibinfo {author} {\bibfnamefont {J.-P.}\ \bibnamefont
  {Adam}}, \bibinfo {author} {\bibfnamefont {Y.}~\bibnamefont {Roussigné}},
  \bibinfo {author} {\bibfnamefont {S.}~\bibnamefont {Eimer}}, \bibinfo
  {author} {\bibfnamefont {T.}~\bibnamefont {Devolder}}, \bibinfo {author}
  {\bibfnamefont {J.-V.}\ \bibnamefont {Kim}}, \bibinfo {author} {\bibfnamefont
  {S.~M.}\ \bibnamefont {Cherif}}, \bibinfo {author} {\bibfnamefont
  {A.}~\bibnamefont {Stashkevich}},\ and\ \bibinfo {author} {\bibfnamefont
  {A.}~\bibnamefont {Thiaville}},\ }\bibfield  {title} {\bibinfo {title}
  {Interfacial {Dzyaloshinskii}-{Moriya} interaction in perpendicularly
  magnetized {Pt}/{Co}/{AlOx} ultrathin films measured by {Brillouin} light
  spectroscopy},\ }\href {https://doi.org/10.1103/PhysRevB.91.180405}
  {\bibfield  {journal} {\bibinfo  {journal} {Physical Review B}\ }\textbf
  {\bibinfo {volume} {91}},\ \bibinfo {pages} {180405} (\bibinfo {year}
  {2015})}\BibitemShut {NoStop}%
\bibitem [{\citenamefont {Thiancourt}\ \emph {et~al.}(2024)\citenamefont
  {Thiancourt}, \citenamefont {Ngom}, \citenamefont {Bardou},\ and\
  \citenamefont {Devolder}}]{thiancourt_unidirectional_2024}%
  \BibitemOpen
  \bibfield  {author} {\bibinfo {author} {\bibfnamefont {G.}~\bibnamefont
  {Thiancourt}}, \bibinfo {author} {\bibfnamefont {S.}~\bibnamefont {Ngom}},
  \bibinfo {author} {\bibfnamefont {N.}~\bibnamefont {Bardou}},\ and\ \bibinfo
  {author} {\bibfnamefont {T.}~\bibnamefont {Devolder}},\ }\bibfield  {title}
  {\bibinfo {title} {Unidirectional spin waves measured using
  propagating-spin-wave spectroscopy},\ }\href
  {https://doi.org/10.1103/PhysRevApplied.22.034040} {\bibfield  {journal}
  {\bibinfo  {journal} {Physical Review Applied}\ }\textbf {\bibinfo {volume}
  {22}},\ \bibinfo {pages} {034040} (\bibinfo {year} {2024})}\BibitemShut
  {NoStop}%
\bibitem [{\citenamefont {Verba}\ \emph {et~al.}(2018)\citenamefont {Verba},
  \citenamefont {Lisenkov}, \citenamefont {Krivorotov}, \citenamefont
  {Tiberkevich},\ and\ \citenamefont {Slavin}}]{verba_nonreciprocal_2018}%
  \BibitemOpen
  \bibfield  {author} {\bibinfo {author} {\bibfnamefont {R.}~\bibnamefont
  {Verba}}, \bibinfo {author} {\bibfnamefont {I.}~\bibnamefont {Lisenkov}},
  \bibinfo {author} {\bibfnamefont {I.}~\bibnamefont {Krivorotov}}, \bibinfo
  {author} {\bibfnamefont {V.}~\bibnamefont {Tiberkevich}},\ and\ \bibinfo
  {author} {\bibfnamefont {A.}~\bibnamefont {Slavin}},\ }\bibfield  {title}
  {\bibinfo {title} {Nonreciprocal {Surface} {Acoustic} {Waves} in
  {Multilayers} with {Magnetoelastic} and {Interfacial}
  {Dzyaloshinskii}-{Moriya} {Interactions}},\ }\href
  {https://doi.org/10.1103/PhysRevApplied.9.064014} {\bibfield  {journal}
  {\bibinfo  {journal} {Physical Review Applied}\ }\textbf {\bibinfo {volume}
  {9}},\ \bibinfo {pages} {064014} (\bibinfo {year} {2018})}\BibitemShut
  {NoStop}%
\bibitem [{\citenamefont {Küß}\ \emph {et~al.}(2024)\citenamefont {Küß},
  \citenamefont {Glamsch}, \citenamefont {Hörner},\ and\ \citenamefont
  {Albrecht}}]{kus_wide-band_2024}%
  \BibitemOpen
  \bibfield  {author} {\bibinfo {author} {\bibfnamefont {M.}~\bibnamefont
  {Küß}}, \bibinfo {author} {\bibfnamefont {S.}~\bibnamefont {Glamsch}},
  \bibinfo {author} {\bibfnamefont {A.}~\bibnamefont {Hörner}},\ and\ \bibinfo
  {author} {\bibfnamefont {M.}~\bibnamefont {Albrecht}},\ }\bibfield  {title}
  {\bibinfo {title} {Wide-{Band} {Nonreciprocal} {Transmission} of {Surface}
  {Acoustic} {Waves} in {Synthetic} {Antiferromagnets}},\ }\href
  {https://doi.org/10.1021/acsaelm.3c01709} {\bibfield  {journal} {\bibinfo
  {journal} {ACS Applied Electronic Materials}\ ,\ \bibinfo {pages}
  {acsaelm.3c01709}} (\bibinfo {year} {2024})}\BibitemShut {NoStop}%
\bibitem [{\citenamefont {Di}\ \emph {et~al.}(2015)\citenamefont {Di},
  \citenamefont {Feng}, \citenamefont {Piramanayagam}, \citenamefont {Zhang},
  \citenamefont {Lim}, \citenamefont {Ng},\ and\ \citenamefont
  {Kuok}}]{di_enhancement_2015}%
  \BibitemOpen
  \bibfield  {author} {\bibinfo {author} {\bibfnamefont {K.}~\bibnamefont
  {Di}}, \bibinfo {author} {\bibfnamefont {S.~X.}\ \bibnamefont {Feng}},
  \bibinfo {author} {\bibfnamefont {S.~N.}\ \bibnamefont {Piramanayagam}},
  \bibinfo {author} {\bibfnamefont {V.~L.}\ \bibnamefont {Zhang}}, \bibinfo
  {author} {\bibfnamefont {H.~S.}\ \bibnamefont {Lim}}, \bibinfo {author}
  {\bibfnamefont {S.~C.}\ \bibnamefont {Ng}},\ and\ \bibinfo {author}
  {\bibfnamefont {M.~H.}\ \bibnamefont {Kuok}},\ }\bibfield  {title} {\bibinfo
  {title} {Enhancement of spin-wave nonreciprocity in magnonic crystals via
  synthetic antiferromagnetic coupling},\ }\href
  {https://doi.org/10.1038/srep10153} {\bibfield  {journal} {\bibinfo
  {journal} {Scientific Reports}\ }\textbf {\bibinfo {volume} {5}},\ \bibinfo
  {pages} {10153} (\bibinfo {year} {2015})}\BibitemShut {NoStop}%
\bibitem [{\citenamefont {Gladii}\ \emph {et~al.}(2016)\citenamefont {Gladii},
  \citenamefont {Haidar}, \citenamefont {Henry}, \citenamefont {Kostylev},\
  and\ \citenamefont {Bailleul}}]{gladii_frequency_2016}%
  \BibitemOpen
  \bibfield  {author} {\bibinfo {author} {\bibfnamefont {O.}~\bibnamefont
  {Gladii}}, \bibinfo {author} {\bibfnamefont {M.}~\bibnamefont {Haidar}},
  \bibinfo {author} {\bibfnamefont {Y.}~\bibnamefont {Henry}}, \bibinfo
  {author} {\bibfnamefont {M.}~\bibnamefont {Kostylev}},\ and\ \bibinfo
  {author} {\bibfnamefont {M.}~\bibnamefont {Bailleul}},\ }\bibfield  {title}
  {\bibinfo {title} {Frequency nonreciprocity of surface spin wave in permalloy
  thin films},\ }\href {https://doi.org/10.1103/PhysRevB.93.054430} {\bibfield
  {journal} {\bibinfo  {journal} {Physical Review B}\ }\textbf {\bibinfo
  {volume} {93}},\ \bibinfo {pages} {054430} (\bibinfo {year} {2016})},\
  \bibinfo {note} {publisher: American Physical Society}\BibitemShut {NoStop}%
\bibitem [{\citenamefont {Gallardo}\ \emph {et~al.}(2019)\citenamefont
  {Gallardo}, \citenamefont {Schneider}, \citenamefont {Chaurasiya},
  \citenamefont {Oelschlägel}, \citenamefont {Arekapudi}, \citenamefont
  {Roldán-Molina}, \citenamefont {Hübner}, \citenamefont {Lenz},
  \citenamefont {Barman}, \citenamefont {Fassbender}, \citenamefont {Lindner},
  \citenamefont {Hellwig},\ and\ \citenamefont
  {Landeros}}]{gallardo_reconfigurable_2019}%
  \BibitemOpen
  \bibfield  {author} {\bibinfo {author} {\bibfnamefont {R.}~\bibnamefont
  {Gallardo}}, \bibinfo {author} {\bibfnamefont {T.}~\bibnamefont {Schneider}},
  \bibinfo {author} {\bibfnamefont {A.}~\bibnamefont {Chaurasiya}}, \bibinfo
  {author} {\bibfnamefont {A.}~\bibnamefont {Oelschlägel}}, \bibinfo {author}
  {\bibfnamefont {S.}~\bibnamefont {Arekapudi}}, \bibinfo {author}
  {\bibfnamefont {A.}~\bibnamefont {Roldán-Molina}}, \bibinfo {author}
  {\bibfnamefont {R.}~\bibnamefont {Hübner}}, \bibinfo {author} {\bibfnamefont
  {K.}~\bibnamefont {Lenz}}, \bibinfo {author} {\bibfnamefont {A.}~\bibnamefont
  {Barman}}, \bibinfo {author} {\bibfnamefont {J.}~\bibnamefont {Fassbender}},
  \bibinfo {author} {\bibfnamefont {J.}~\bibnamefont {Lindner}}, \bibinfo
  {author} {\bibfnamefont {O.}~\bibnamefont {Hellwig}},\ and\ \bibinfo {author}
  {\bibfnamefont {P.}~\bibnamefont {Landeros}},\ }\bibfield  {title} {\bibinfo
  {title} {Reconfigurable {Spin}-{Wave} {Nonreciprocity} {Induced} by {Dipolar}
  {Interaction} in a {Coupled} {Ferromagnetic} {Bilayer}},\ }\href
  {https://doi.org/10.1103/PhysRevApplied.12.034012} {\bibfield  {journal}
  {\bibinfo  {journal} {Physical Review Applied}\ }\textbf {\bibinfo {volume}
  {12}},\ \bibinfo {pages} {034012} (\bibinfo {year} {2019})}\BibitemShut
  {NoStop}%
\bibitem [{\citenamefont {Gallardo}\ \emph {et~al.}(2021)\citenamefont
  {Gallardo}, \citenamefont {Alvarado-Seguel}, \citenamefont {Kákay},
  \citenamefont {Lindner},\ and\ \citenamefont
  {Landeros}}]{gallardo_spin-wave_2021}%
  \BibitemOpen
  \bibfield  {author} {\bibinfo {author} {\bibfnamefont {R.~A.}\ \bibnamefont
  {Gallardo}}, \bibinfo {author} {\bibfnamefont {P.}~\bibnamefont
  {Alvarado-Seguel}}, \bibinfo {author} {\bibfnamefont {A.}~\bibnamefont
  {Kákay}}, \bibinfo {author} {\bibfnamefont {J.}~\bibnamefont {Lindner}},\
  and\ \bibinfo {author} {\bibfnamefont {P.}~\bibnamefont {Landeros}},\
  }\bibfield  {title} {\bibinfo {title} {Spin-wave focusing induced by
  dipole-dipole interaction in synthetic antiferromagnets},\ }\href
  {https://doi.org/10.1103/PhysRevB.104.174417} {\bibfield  {journal} {\bibinfo
   {journal} {Physical Review B}\ }\textbf {\bibinfo {volume} {104}},\ \bibinfo
  {pages} {174417} (\bibinfo {year} {2021})}\BibitemShut {NoStop}%
\bibitem [{\citenamefont {Matsumoto}\ \emph {et~al.}(2022)\citenamefont
  {Matsumoto}, \citenamefont {Kawada}, \citenamefont {Ishibashi}, \citenamefont
  {Kawaguchi},\ and\ \citenamefont {Hayashi}}]{matsumoto_large_2022}%
  \BibitemOpen
  \bibfield  {author} {\bibinfo {author} {\bibfnamefont {H.}~\bibnamefont
  {Matsumoto}}, \bibinfo {author} {\bibfnamefont {T.}~\bibnamefont {Kawada}},
  \bibinfo {author} {\bibfnamefont {M.}~\bibnamefont {Ishibashi}}, \bibinfo
  {author} {\bibfnamefont {M.}~\bibnamefont {Kawaguchi}},\ and\ \bibinfo
  {author} {\bibfnamefont {M.}~\bibnamefont {Hayashi}},\ }\bibfield  {title}
  {\bibinfo {title} {Large surface acoustic wave nonreciprocity in synthetic
  antiferromagnets},\ }\href
  {https://iopscience.iop.org/article/10.35848/1882-0786/ac6da1/meta}
  {\bibfield  {journal} {\bibinfo  {journal} {Applied Physics Express}\
  }\textbf {\bibinfo {volume} {15}},\ \bibinfo {pages} {063003} (\bibinfo
  {year} {2022})}\BibitemShut {NoStop}%
\bibitem [{\citenamefont {Mruczkiewicz}\ and\ \citenamefont
  {Krawczyk}(2016)}]{Mruczkiewicz_influence_2016}%
  \BibitemOpen
  \bibfield  {author} {\bibinfo {author} {\bibfnamefont {M.}~\bibnamefont
  {Mruczkiewicz}}\ and\ \bibinfo {author} {\bibfnamefont {M.}~\bibnamefont
  {Krawczyk}},\ }\bibfield  {title} {\bibinfo {title} {Influence of the
  {Dzyaloshinskii}-{Moriya} interaction on the {FMR} spectrum of magnonic
  crystals and confined structures},\ }\href
  {https://doi.org/10.1103/PhysRevB.94.024434} {\bibfield  {journal} {\bibinfo
  {journal} {Physical Review B}\ }\textbf {\bibinfo {volume} {94}},\ \bibinfo
  {pages} {024434} (\bibinfo {year} {2016})}\BibitemShut {NoStop}%
\bibitem [{\citenamefont {Zingsem}\ \emph {et~al.}(2019)\citenamefont
  {Zingsem}, \citenamefont {Farle}, \citenamefont {Stamps},\ and\ \citenamefont
  {Camley}}]{zingsem_unusual_2019}%
  \BibitemOpen
  \bibfield  {author} {\bibinfo {author} {\bibfnamefont {B.~W.}\ \bibnamefont
  {Zingsem}}, \bibinfo {author} {\bibfnamefont {M.}~\bibnamefont {Farle}},
  \bibinfo {author} {\bibfnamefont {R.~L.}\ \bibnamefont {Stamps}},\ and\
  \bibinfo {author} {\bibfnamefont {R.~E.}\ \bibnamefont {Camley}},\ }\bibfield
   {title} {\bibinfo {title} {Unusual nature of confined modes in a chiral
  system: {Directional} transport in standing waves},\ }\href
  {https://doi.org/10.1103/PhysRevB.99.214429} {\bibfield  {journal} {\bibinfo
  {journal} {Physical Review B}\ }\textbf {\bibinfo {volume} {99}},\ \bibinfo
  {pages} {214429} (\bibinfo {year} {2019})}\BibitemShut {NoStop}%
\bibitem [{\citenamefont {Silvani}\ \emph {et~al.}(2021)\citenamefont
  {Silvani}, \citenamefont {Alunni}, \citenamefont {Tacchi},\ and\
  \citenamefont {Carlotti}}]{silvani_effect_2021}%
  \BibitemOpen
  \bibfield  {author} {\bibinfo {author} {\bibfnamefont {R.}~\bibnamefont
  {Silvani}}, \bibinfo {author} {\bibfnamefont {M.}~\bibnamefont {Alunni}},
  \bibinfo {author} {\bibfnamefont {S.}~\bibnamefont {Tacchi}},\ and\ \bibinfo
  {author} {\bibfnamefont {G.}~\bibnamefont {Carlotti}},\ }\bibfield  {title}
  {\bibinfo {title} {Effect of the {Interfacial} {Dzyaloshinskii}–{Moriya}
  {Interaction} on the {Spin} {Waves} {Eigenmodes} of {Isolated} {Stripes} and
  {Dots} {Magnetized} {In}-{Plane}: {A} {Micromagnetic} {Study}},\ }\href
  {https://doi.org/10.3390/app11072929} {\bibfield  {journal} {\bibinfo
  {journal} {Applied Sciences}\ }\textbf {\bibinfo {volume} {11}},\ \bibinfo
  {pages} {2929} (\bibinfo {year} {2021})}\BibitemShut {NoStop}%
\bibitem [{\citenamefont {Tacchi}\ \emph {et~al.}(2023)\citenamefont {Tacchi},
  \citenamefont {Silvani}, \citenamefont {Kuepferling}, \citenamefont
  {Fernández~Scarioni}, \citenamefont {Sievers}, \citenamefont {Schumacher},
  \citenamefont {Darwin}, \citenamefont {Syskaki}, \citenamefont {Jakob},
  \citenamefont {Kläui},\ and\ \citenamefont
  {Carlotti}}]{tacchi_suppression_2023}%
  \BibitemOpen
  \bibfield  {author} {\bibinfo {author} {\bibfnamefont {S.}~\bibnamefont
  {Tacchi}}, \bibinfo {author} {\bibfnamefont {R.}~\bibnamefont {Silvani}},
  \bibinfo {author} {\bibfnamefont {M.}~\bibnamefont {Kuepferling}}, \bibinfo
  {author} {\bibfnamefont {A.}~\bibnamefont {Fernández~Scarioni}}, \bibinfo
  {author} {\bibfnamefont {S.}~\bibnamefont {Sievers}}, \bibinfo {author}
  {\bibfnamefont {H.~W.}\ \bibnamefont {Schumacher}}, \bibinfo {author}
  {\bibfnamefont {E.}~\bibnamefont {Darwin}}, \bibinfo {author} {\bibfnamefont
  {M.-A.}\ \bibnamefont {Syskaki}}, \bibinfo {author} {\bibfnamefont
  {G.}~\bibnamefont {Jakob}}, \bibinfo {author} {\bibfnamefont
  {M.}~\bibnamefont {Kläui}},\ and\ \bibinfo {author} {\bibfnamefont
  {G.}~\bibnamefont {Carlotti}},\ }\bibfield  {title} {\bibinfo {title}
  {Suppression of spin-wave nonreciprocity due to interfacial
  {Dzyaloshinskii}-{Moriya} interaction by lateral confinement in magnetic
  nanostructures},\ }\href {https://doi.org/10.1103/PhysRevB.108.024430}
  {\bibfield  {journal} {\bibinfo  {journal} {Physical Review B}\ }\textbf
  {\bibinfo {volume} {108}},\ \bibinfo {pages} {024430} (\bibinfo {year}
  {2023})}\BibitemShut {NoStop}%
\bibitem [{\citenamefont {Makartsou}\ \emph {et~al.}(2024)\citenamefont
  {Makartsou}, \citenamefont {Gołebiewski}, \citenamefont {Guzowska},
  \citenamefont {Stognij}, \citenamefont {Gieniusz},\ and\ \citenamefont
  {Krawczyk}}]{makartsou_spin-wave_2024}%
  \BibitemOpen
  \bibfield  {author} {\bibinfo {author} {\bibfnamefont {U.}~\bibnamefont
  {Makartsou}}, \bibinfo {author} {\bibfnamefont {M.}~\bibnamefont
  {Golebiewski}}, \bibinfo {author} {\bibfnamefont {U.}~\bibnamefont
  {Guzowska}}, \bibinfo {author} {\bibfnamefont {A.}~\bibnamefont {Stognij}},
  \bibinfo {author} {\bibfnamefont {R.}~\bibnamefont {Gieniusz}},\ and\
  \bibinfo {author} {\bibfnamefont {M.}~\bibnamefont {Krawczyk}},\ }\bibfield
  {title} {\bibinfo {title} {Spin-wave self-imaging: {Experimental} and
  numerical demonstration of caustic and {Talbot}-like diffraction patterns},\
  }\href {https://doi.org/10.1063/5.0195099} {\bibfield  {journal} {\bibinfo
  {journal} {Applied Physics Letters}\ }\textbf {\bibinfo {volume} {124}},\
  \bibinfo {pages} {192406} (\bibinfo {year} {2024})}\BibitemShut {NoStop}%
\bibitem [{\citenamefont {Körner}\ \emph {et~al.}(2017)\citenamefont
  {Körner}, \citenamefont {Stigloher},\ and\ \citenamefont
  {Back}}]{korner_excitation_2017}%
  \BibitemOpen
  \bibfield  {author} {\bibinfo {author} {\bibfnamefont {H.~S.}\ \bibnamefont
  {Körner}}, \bibinfo {author} {\bibfnamefont {J.}~\bibnamefont {Stigloher}},\
  and\ \bibinfo {author} {\bibfnamefont {C.~H.}\ \bibnamefont {Back}},\
  }\bibfield  {title} {\bibinfo {title} {Excitation and tailoring of
  diffractive spin-wave beams in {NiFe} using nonuniform microwave antennas},\
  }\href {https://doi.org/10.1103/PhysRevB.96.100401} {\bibfield  {journal}
  {\bibinfo  {journal} {Physical Review B}\ }\textbf {\bibinfo {volume} {96}},\
  \bibinfo {pages} {100401} (\bibinfo {year} {2017})},\ \bibinfo {note}
  {publisher: American Physical Society}\BibitemShut {NoStop}%
\bibitem [{\citenamefont {Lock}(2008)}]{lock_properties_2008}%
  \BibitemOpen
  \bibfield  {author} {\bibinfo {author} {\bibfnamefont {E.~H.}\ \bibnamefont
  {Lock}},\ }\bibfield  {title} {\bibinfo {title} {The properties of
  isofrequency dependences and the laws of geometrical optics},\ }\href
  {https://doi.org/10.1070/PU2008v051n04ABEH006460} {\bibfield  {journal}
  {\bibinfo  {journal} {Physics-Uspekhi}\ }\textbf {\bibinfo {volume} {51}},\
  \bibinfo {pages} {375} (\bibinfo {year} {2008})},\ \bibinfo {note}
  {publisher: IOP Publishing}\BibitemShut {NoStop}%
\bibitem [{\citenamefont {Bailleul}\ \emph {et~al.}(2003)\citenamefont
  {Bailleul}, \citenamefont {Olligs},\ and\ \citenamefont
  {Fermon}}]{bailleul_propagating_2003}%
  \BibitemOpen
  \bibfield  {author} {\bibinfo {author} {\bibfnamefont {M.}~\bibnamefont
  {Bailleul}}, \bibinfo {author} {\bibfnamefont {D.}~\bibnamefont {Olligs}},\
  and\ \bibinfo {author} {\bibfnamefont {C.}~\bibnamefont {Fermon}},\
  }\bibfield  {title} {\bibinfo {title} {Propagating spin wave spectroscopy in
  a permalloy film: {A} quantitative analysis},\ }\href
  {https://doi.org/10.1063/1.1597745} {\bibfield  {journal} {\bibinfo
  {journal} {Applied Physics Letters}\ }\textbf {\bibinfo {volume} {83}},\
  \bibinfo {pages} {972} (\bibinfo {year} {2003})}\BibitemShut {NoStop}%
\bibitem [{\citenamefont {Demidov}\ \emph {et~al.}(2008)\citenamefont
  {Demidov}, \citenamefont {Demokritov}, \citenamefont {Rott}, \citenamefont
  {Krzysteczko},\ and\ \citenamefont {Reiss}}]{demidov_mode_2008}%
  \BibitemOpen
  \bibfield  {author} {\bibinfo {author} {\bibfnamefont {V.~E.}\ \bibnamefont
  {Demidov}}, \bibinfo {author} {\bibfnamefont {S.~O.}\ \bibnamefont
  {Demokritov}}, \bibinfo {author} {\bibfnamefont {K.}~\bibnamefont {Rott}},
  \bibinfo {author} {\bibfnamefont {P.}~\bibnamefont {Krzysteczko}},\ and\
  \bibinfo {author} {\bibfnamefont {G.}~\bibnamefont {Reiss}},\ }\bibfield
  {title} {\bibinfo {title} {Mode interference and periodic self-focusing of
  spin waves in permalloy microstripes},\ }\href
  {https://doi.org/10.1103/PhysRevB.77.064406} {\bibfield  {journal} {\bibinfo
  {journal} {Physical Review B}\ }\textbf {\bibinfo {volume} {77}},\ \bibinfo
  {pages} {064406} (\bibinfo {year} {2008})}\BibitemShut {NoStop}%
\bibitem [{\citenamefont
  {Devolder}(2023)}]{devolder_propagating-spin-wave_2023}%
  \BibitemOpen
  \bibfield  {author} {\bibinfo {author} {\bibfnamefont {T.}~\bibnamefont
  {Devolder}},\ }\bibfield  {title} {\bibinfo {title} {Propagating-spin-wave
  spectroscopy using inductive antennas: {Conditions} for unidirectional energy
  flow},\ }\href {https://doi.org/10.1103/PhysRevApplied.20.054057} {\bibfield
  {journal} {\bibinfo  {journal} {Physical Review Applied}\ }\textbf {\bibinfo
  {volume} {20}},\ \bibinfo {pages} {054057} (\bibinfo {year}
  {2023})}\BibitemShut {NoStop}%
\bibitem [{\citenamefont {Pirro}\ \emph {et~al.}(2014)\citenamefont {Pirro},
  \citenamefont {Brächer}, \citenamefont {Chumak}, \citenamefont {Lägel},
  \citenamefont {Dubs}, \citenamefont {Surzhenko}, \citenamefont {Görnert},
  \citenamefont {Leven},\ and\ \citenamefont
  {Hillebrands}}]{pirro_spin-wave_2014}%
  \BibitemOpen
  \bibfield  {author} {\bibinfo {author} {\bibfnamefont {P.}~\bibnamefont
  {Pirro}}, \bibinfo {author} {\bibfnamefont {T.}~\bibnamefont {Brächer}},
  \bibinfo {author} {\bibfnamefont {A.~V.}\ \bibnamefont {Chumak}}, \bibinfo
  {author} {\bibfnamefont {B.}~\bibnamefont {Lägel}}, \bibinfo {author}
  {\bibfnamefont {C.}~\bibnamefont {Dubs}}, \bibinfo {author} {\bibfnamefont
  {O.}~\bibnamefont {Surzhenko}}, \bibinfo {author} {\bibfnamefont
  {P.}~\bibnamefont {Görnert}}, \bibinfo {author} {\bibfnamefont
  {B.}~\bibnamefont {Leven}},\ and\ \bibinfo {author} {\bibfnamefont
  {B.}~\bibnamefont {Hillebrands}},\ }\bibfield  {title} {\bibinfo {title}
  {Spin-wave excitation and propagation in microstructured waveguides of
  yttrium iron garnet/{Pt} bilayers},\ }\href
  {https://doi.org/10.1063/1.4861343} {\bibfield  {journal} {\bibinfo
  {journal} {Applied Physics Letters}\ }\textbf {\bibinfo {volume} {104}},\
  \bibinfo {pages} {012402} (\bibinfo {year} {2014})}\BibitemShut {NoStop}%
\bibitem [{\citenamefont {Demidov}\ \emph {et~al.}(2016)\citenamefont
  {Demidov}, \citenamefont {Urazhdin}, \citenamefont {Liu}, \citenamefont
  {Divinskiy}, \citenamefont {Telegin},\ and\ \citenamefont
  {Demokritov}}]{demidov_excitation_2016}%
  \BibitemOpen
  \bibfield  {author} {\bibinfo {author} {\bibfnamefont {V.~E.}\ \bibnamefont
  {Demidov}}, \bibinfo {author} {\bibfnamefont {S.}~\bibnamefont {Urazhdin}},
  \bibinfo {author} {\bibfnamefont {R.}~\bibnamefont {Liu}}, \bibinfo {author}
  {\bibfnamefont {B.}~\bibnamefont {Divinskiy}}, \bibinfo {author}
  {\bibfnamefont {A.}~\bibnamefont {Telegin}},\ and\ \bibinfo {author}
  {\bibfnamefont {S.~O.}\ \bibnamefont {Demokritov}},\ }\bibfield  {title}
  {\bibinfo {title} {Excitation of coherent propagating spin waves by pure spin
  currents},\ }\href {https://doi.org/10.1038/ncomm50446} {\bibfield
  {journal} {\bibinfo  {journal} {Nature Communications}\ }\textbf {\bibinfo
  {volume} {7}},\ \bibinfo {pages} {1} (\bibinfo {year} {2016})}\BibitemShut
  {NoStop}%
\bibitem [{\citenamefont {Bayer}\ \emph {et~al.}(2006)\citenamefont {Bayer},
  \citenamefont {Jorzick}, \citenamefont {Demokritov}, \citenamefont {Slavin},
  \citenamefont {Guslienko}, \citenamefont {Berkov}, \citenamefont {Gorn},
  \citenamefont {Kostylev},\ and\ \citenamefont
  {Hillebrands}}]{bayer_spin-wave_2006}%
  \BibitemOpen
  \bibfield  {author} {\bibinfo {author} {\bibfnamefont {C.}~\bibnamefont
  {Bayer}}, \bibinfo {author} {\bibfnamefont {J.}~\bibnamefont {Jorzick}},
  \bibinfo {author} {\bibfnamefont {S.~O.}\ \bibnamefont {Demokritov}},
  \bibinfo {author} {\bibfnamefont {A.~N.}\ \bibnamefont {Slavin}}, \bibinfo
  {author} {\bibfnamefont {K.~Y.}\ \bibnamefont {Guslienko}}, \bibinfo {author}
  {\bibfnamefont {D.~V.}\ \bibnamefont {Berkov}}, \bibinfo {author}
  {\bibfnamefont {N.~L.}\ \bibnamefont {Gorn}}, \bibinfo {author}
  {\bibfnamefont {M.~P.}\ \bibnamefont {Kostylev}},\ and\ \bibinfo {author}
  {\bibfnamefont {B.}~\bibnamefont {Hillebrands}},\ }\bibfield  {title}
  {\bibinfo {title} {Spin-{Wave} {Excitations} in {Finite} {Rectangular}
  {Elements}},\ }in\ \href {https://doi.org/10.1007/10938171_2} {\emph
  {\bibinfo {booktitle} {Spin {Dynamics} in {Confined} {Magnetic} {Structures}
  {III}}}},\ \bibinfo {editor} {edited by\ \bibinfo {editor} {\bibfnamefont
  {B.}~\bibnamefont {Hillebrands}}\ and\ \bibinfo {editor} {\bibfnamefont
  {A.}~\bibnamefont {Thiaville}}}\ (\bibinfo  {publisher} {Springer},\ \bibinfo
  {address} {Berlin, Heidelberg},\ \bibinfo {year} {2006})\ pp.\ \bibinfo
  {pages} {57--103}\BibitemShut {NoStop}%
\bibitem [{\citenamefont {Guslienko}\ \emph {et~al.}(2002)\citenamefont
  {Guslienko}, \citenamefont {Demokritov}, \citenamefont {Hillebrands},\ and\
  \citenamefont {Slavin}}]{guslienko_effective_2002}%
  \BibitemOpen
  \bibfield  {author} {\bibinfo {author} {\bibfnamefont {K.~Y.}\ \bibnamefont
  {Guslienko}}, \bibinfo {author} {\bibfnamefont {S.~O.}\ \bibnamefont
  {Demokritov}}, \bibinfo {author} {\bibfnamefont {B.}~\bibnamefont
  {Hillebrands}},\ and\ \bibinfo {author} {\bibfnamefont {A.~N.}\ \bibnamefont
  {Slavin}},\ }\bibfield  {title} {\bibinfo {title} {Effective dipolar boundary
  conditions for dynamic magnetization in thin magnetic stripes},\ }\href
  {https://doi.org/10.1103/PhysRevB.66.132402} {\bibfield  {journal} {\bibinfo
  {journal} {Physical Review B}\ }\textbf {\bibinfo {volume} {66}},\ \bibinfo
  {pages} {132402} (\bibinfo {year} {2002})}\BibitemShut {NoStop}%
\bibitem [{\citenamefont {Demidov}\ and\ \citenamefont
  {Demokritov}(2015)}]{demidov_magnonic_2015}%
  \BibitemOpen
  \bibfield  {author} {\bibinfo {author} {\bibfnamefont {V.~E.}\ \bibnamefont
  {Demidov}}\ and\ \bibinfo {author} {\bibfnamefont {S.~O.}\ \bibnamefont
  {Demokritov}},\ }\bibfield  {title} {\bibinfo {title} {Magnonic {Waveguides}
  {Studied} by {Microfocus} {Brillouin} {Light} {Scattering}},\ }\href
  {https://doi.org/10.1109/TMAG.2014.2388196} {\bibfield  {journal} {\bibinfo
  {journal} {IEEE Transactions on Magnetics}\ }\textbf {\bibinfo {volume}
  {51}},\ \bibinfo {pages} {1} (\bibinfo {year} {2015})}\BibitemShut {NoStop}%
\bibitem [{\citenamefont {Talmelli}\ \emph {et~al.}(2020)\citenamefont
  {Talmelli}, \citenamefont {Devolder}, \citenamefont {Träger}, \citenamefont
  {Förster}, \citenamefont {Wintz}, \citenamefont {Weigand}, \citenamefont
  {Stoll}, \citenamefont {Heyns}, \citenamefont {Schütz}, \citenamefont
  {Radu}, \citenamefont {Gräfe}, \citenamefont {Ciubotaru},\ and\
  \citenamefont {Adelmann}}]{talmelli_reconfigurable_2020}%
  \BibitemOpen
  \bibfield  {author} {\bibinfo {author} {\bibfnamefont {G.}~\bibnamefont
  {Talmelli}}, \bibinfo {author} {\bibfnamefont {T.}~\bibnamefont {Devolder}},
  \bibinfo {author} {\bibfnamefont {N.}~\bibnamefont {Träger}}, \bibinfo
  {author} {\bibfnamefont {J.}~\bibnamefont {Förster}}, \bibinfo {author}
  {\bibfnamefont {S.}~\bibnamefont {Wintz}}, \bibinfo {author} {\bibfnamefont
  {M.}~\bibnamefont {Weigand}}, \bibinfo {author} {\bibfnamefont
  {H.}~\bibnamefont {Stoll}}, \bibinfo {author} {\bibfnamefont
  {M.}~\bibnamefont {Heyns}}, \bibinfo {author} {\bibfnamefont
  {G.}~\bibnamefont {Schütz}}, \bibinfo {author} {\bibfnamefont {I.~P.}\
  \bibnamefont {Radu}}, \bibinfo {author} {\bibfnamefont {J.}~\bibnamefont
  {Gräfe}}, \bibinfo {author} {\bibfnamefont {F.}~\bibnamefont {Ciubotaru}},\
  and\ \bibinfo {author} {\bibfnamefont {C.}~\bibnamefont {Adelmann}},\
  }\bibfield  {title} {\bibinfo {title} {Reconfigurable submicrometer spin-wave
  majority gate with electrical transducers},\ }\href
  {https://doi.org/10.1126/sciadv.abb4042} {\bibfield  {journal} {\bibinfo
  {journal} {Science Advances}\ }\textbf {\bibinfo {volume} {6}},\ \bibinfo
  {pages} {eabb4042} (\bibinfo {year} {2020})}\BibitemShut {NoStop}%
\bibitem [{\citenamefont {Shiota}\ \emph {et~al.}(2020)\citenamefont {Shiota},
  \citenamefont {Taniguchi}, \citenamefont {Ishibashi}, \citenamefont
  {Moriyama},\ and\ \citenamefont {Ono}}]{shiota_tunable_2020}%
  \BibitemOpen
  \bibfield  {author} {\bibinfo {author} {\bibfnamefont {Y.}~\bibnamefont
  {Shiota}}, \bibinfo {author} {\bibfnamefont {T.}~\bibnamefont {Taniguchi}},
  \bibinfo {author} {\bibfnamefont {M.}~\bibnamefont {Ishibashi}}, \bibinfo
  {author} {\bibfnamefont {T.}~\bibnamefont {Moriyama}},\ and\ \bibinfo
  {author} {\bibfnamefont {T.}~\bibnamefont {Ono}},\ }\bibfield  {title}
  {\bibinfo {title} {Tunable {Magnon}-{Magnon} {Coupling} {Mediated} by
  {Dynamic} {Dipolar} {Interaction} in {Synthetic} {Antiferromagnets}},\ }\href
  {https://doi.org/10.1103/PhysRevLett.125.017203} {\bibfield  {journal}
  {\bibinfo  {journal} {Physical Review Letters}\ }\textbf {\bibinfo {volume}
  {125}},\ \bibinfo {pages} {017203} (\bibinfo {year} {2020})}\BibitemShut
  {NoStop}%
\bibitem [{\citenamefont {Mouhoub}\ \emph {et~al.}(2023)\citenamefont
  {Mouhoub}, \citenamefont {Millo}, \citenamefont {Chappert}, \citenamefont
  {Kim}, \citenamefont {Létang}, \citenamefont {Solignac},\ and\ \citenamefont
  {Devolder}}]{mouhoub_exchange_2023}%
  \BibitemOpen
  \bibfield  {author} {\bibinfo {author} {\bibfnamefont {A.}~\bibnamefont
  {Mouhoub}}, \bibinfo {author} {\bibfnamefont {F.}~\bibnamefont {Millo}},
  \bibinfo {author} {\bibfnamefont {C.}~\bibnamefont {Chappert}}, \bibinfo
  {author} {\bibfnamefont {J.-V.}\ \bibnamefont {Kim}}, \bibinfo {author}
  {\bibfnamefont {J.}~\bibnamefont {Létang}}, \bibinfo {author} {\bibfnamefont
  {A.}~\bibnamefont {Solignac}},\ and\ \bibinfo {author} {\bibfnamefont
  {T.}~\bibnamefont {Devolder}},\ }\bibfield  {title} {\bibinfo {title}
  {Exchange energies in {CoFeB}/{Ru}/{CoFeB} synthetic antiferromagnets},\
  }\href {https://doi.org/10.1103/PhysRevMaterials.7.044404} {\bibfield
  {journal} {\bibinfo  {journal} {Physical Review Materials}\ }\textbf
  {\bibinfo {volume} {7}},\ \bibinfo {pages} {044404} (\bibinfo {year}
  {2023})}\BibitemShut {NoStop}%
\bibitem [{\citenamefont {Vansteenkiste}\ \emph {et~al.}(2014)\citenamefont
  {Vansteenkiste}, \citenamefont {Leliaert}, \citenamefont {Dvornik},
  \citenamefont {Helsen}, \citenamefont {Garcia-Sanchez},\ and\ \citenamefont
  {Waeyenberge}}]{vansteenkiste_design_2014}%
  \BibitemOpen
  \bibfield  {author} {\bibinfo {author} {\bibfnamefont {A.}~\bibnamefont
  {Vansteenkiste}}, \bibinfo {author} {\bibfnamefont {J.}~\bibnamefont
  {Leliaert}}, \bibinfo {author} {\bibfnamefont {M.}~\bibnamefont {Dvornik}},
  \bibinfo {author} {\bibfnamefont {M.}~\bibnamefont {Helsen}}, \bibinfo
  {author} {\bibfnamefont {F.}~\bibnamefont {Garcia-Sanchez}},\ and\ \bibinfo
  {author} {\bibfnamefont {B.~V.}\ \bibnamefont {Waeyenberge}},\ }\bibfield
  {title} {\bibinfo {title} {The design and verification of {MuMax3}},\ }\href
  {https://doi.org/10.1063/1.4899186} {\bibfield  {journal} {\bibinfo
  {journal} {AIP Advances}\ }\textbf {\bibinfo {volume} {4}},\ \bibinfo {pages}
  {107133} (\bibinfo {year} {2014})}\BibitemShut {NoStop}%
\bibitem [{\citenamefont {Hamrle}\ \emph {et~al.}(2010)\citenamefont {Hamrle},
  \citenamefont {Pištora}, \citenamefont {Hillebrands}, \citenamefont {Lenk},\
  and\ \citenamefont {Münzenberg}}]{hamrle_analytical_2010}%
  \BibitemOpen
  \bibfield  {author} {\bibinfo {author} {\bibfnamefont {J.}~\bibnamefont
  {Hamrle}}, \bibinfo {author} {\bibfnamefont {J.}~\bibnamefont {Pištora}},
  \bibinfo {author} {\bibfnamefont {B.}~\bibnamefont {Hillebrands}}, \bibinfo
  {author} {\bibfnamefont {B.}~\bibnamefont {Lenk}},\ and\ \bibinfo {author}
  {\bibfnamefont {M.}~\bibnamefont {Münzenberg}},\ }\bibfield  {title}
  {\bibinfo {title} {Analytical expression of the magneto-optical {Kerr} effect
  and {Brillouin} light scattering intensity arising from dynamic
  magnetization},\ }\href {https://doi.org/10.1088/0022-3727/43/32/325004}
  {\bibfield  {journal} {\bibinfo  {journal} {Journal of Physics D: Applied
  Physics}\ }\textbf {\bibinfo {volume} {43}},\ \bibinfo {pages} {325004}
  (\bibinfo {year} {2010})}\BibitemShut {NoStop}%
\bibitem [{\citenamefont {Nörtemann}\ \emph {et~al.}(1993)\citenamefont
  {Nörtemann}, \citenamefont {Stamps},\ and\ \citenamefont
  {Camley}}]{Nortemann_microscopic_1993}%
  \BibitemOpen
  \bibfield  {author} {\bibinfo {author} {\bibfnamefont {F.~C.}\ \bibnamefont
  {Nörtemann}}, \bibinfo {author} {\bibfnamefont {R.~L.}\ \bibnamefont
  {Stamps}},\ and\ \bibinfo {author} {\bibfnamefont {R.~E.}\ \bibnamefont
  {Camley}},\ }\bibfield  {title} {\bibinfo {title} {Microscopic calculation of
  spin waves in antiferromagnetically coupled multilayers: {Nonreciprocity} and
  finite-size effects},\ }\href {https://doi.org/10.1103/PhysRevB.47.11910}
  {\bibfield  {journal} {\bibinfo  {journal} {Physical Review B}\ }\textbf
  {\bibinfo {volume} {47}},\ \bibinfo {pages} {11910} (\bibinfo {year}
  {1993})}\BibitemShut {NoStop}%
\bibitem [{\citenamefont {Grimsditch}\ \emph {et~al.}(2004)\citenamefont
  {Grimsditch}, \citenamefont {Giovannini}, \citenamefont {Montoncello},
  \citenamefont {Nizzoli}, \citenamefont {Leaf},\ and\ \citenamefont
  {Kaper}}]{Grimsditch_magnetic_2004}%
  \BibitemOpen
  \bibfield  {author} {\bibinfo {author} {\bibfnamefont {M.}~\bibnamefont
  {Grimsditch}}, \bibinfo {author} {\bibfnamefont {L.}~\bibnamefont
  {Giovannini}}, \bibinfo {author} {\bibfnamefont {F.}~\bibnamefont
  {Montoncello}}, \bibinfo {author} {\bibfnamefont {F.}~\bibnamefont
  {Nizzoli}}, \bibinfo {author} {\bibfnamefont {G.~K.}\ \bibnamefont {Leaf}},\
  and\ \bibinfo {author} {\bibfnamefont {H.~G.}\ \bibnamefont {Kaper}},\
  }\bibfield  {title} {\bibinfo {title} {Magnetic normal modes in ferromagnetic
  nanoparticles: {A} dynamical matrix approach},\ }\href
  {https://doi.org/10.1103/PhysRevB.70.054409} {\bibfield  {journal} {\bibinfo
  {journal} {Physical Review B}\ }\textbf {\bibinfo {volume} {70}},\ \bibinfo
  {pages} {054409} (\bibinfo {year} {2004})}\BibitemShut {NoStop}%
\bibitem [{\citenamefont {Henry}\ \emph {et~al.}(2016)\citenamefont {Henry},
  \citenamefont {Gladii},\ and\ \citenamefont
  {Bailleul}}]{henry_propagating_2016}%
  \BibitemOpen
  \bibfield  {author} {\bibinfo {author} {\bibfnamefont {Y.}~\bibnamefont
  {Henry}}, \bibinfo {author} {\bibfnamefont {O.}~\bibnamefont {Gladii}},\ and\
  \bibinfo {author} {\bibfnamefont {M.}~\bibnamefont {Bailleul}},\ }\bibfield
  {title} {\bibinfo {title} {Propagating spin-wave normal modes: {A} dynamic
  matrix approach using plane-wave demagnetizating tensors},\ }\href
  {http://arxiv.org/abs/1611.06153} {\bibfield  {journal} {\bibinfo  {journal}
  {arXiv:1611.06153 [cond-mat]}\ } (\bibinfo {year} {2016})},\ \bibinfo {note}
  {arXiv: 1611.06153}\BibitemShut {NoStop}%
\bibitem [{\citenamefont {Millo}\ \emph {et~al.}(2023)\citenamefont {Millo},
  \citenamefont {Adam}, \citenamefont {Chappert}, \citenamefont {Kim},
  \citenamefont {Mouhoub}, \citenamefont {Solignac},\ and\ \citenamefont
  {Devolder}}]{millo_unidirectionality_2023}%
  \BibitemOpen
  \bibfield  {author} {\bibinfo {author} {\bibfnamefont {F.}~\bibnamefont
  {Millo}}, \bibinfo {author} {\bibfnamefont {J.-P.}\ \bibnamefont {Adam}},
  \bibinfo {author} {\bibfnamefont {C.}~\bibnamefont {Chappert}}, \bibinfo
  {author} {\bibfnamefont {J.-V.}\ \bibnamefont {Kim}}, \bibinfo {author}
  {\bibfnamefont {A.}~\bibnamefont {Mouhoub}}, \bibinfo {author} {\bibfnamefont
  {A.}~\bibnamefont {Solignac}},\ and\ \bibinfo {author} {\bibfnamefont
  {T.}~\bibnamefont {Devolder}},\ }\bibfield  {title} {\bibinfo {title}
  {Unidirectionality of spin waves in synthetic antiferromagnets},\ }\href
  {https://doi.org/10.1103/PhysRevApplied.20.054051} {\bibfield  {journal}
  {\bibinfo  {journal} {Physical Review Applied}\ }\textbf {\bibinfo {volume}
  {20}},\ \bibinfo {pages} {054051} (\bibinfo {year} {2023})}\BibitemShut
  {NoStop}%
\end{thebibliography}
%

\newpage


%
\begin{figure}
\includegraphics[width=\linewidth]{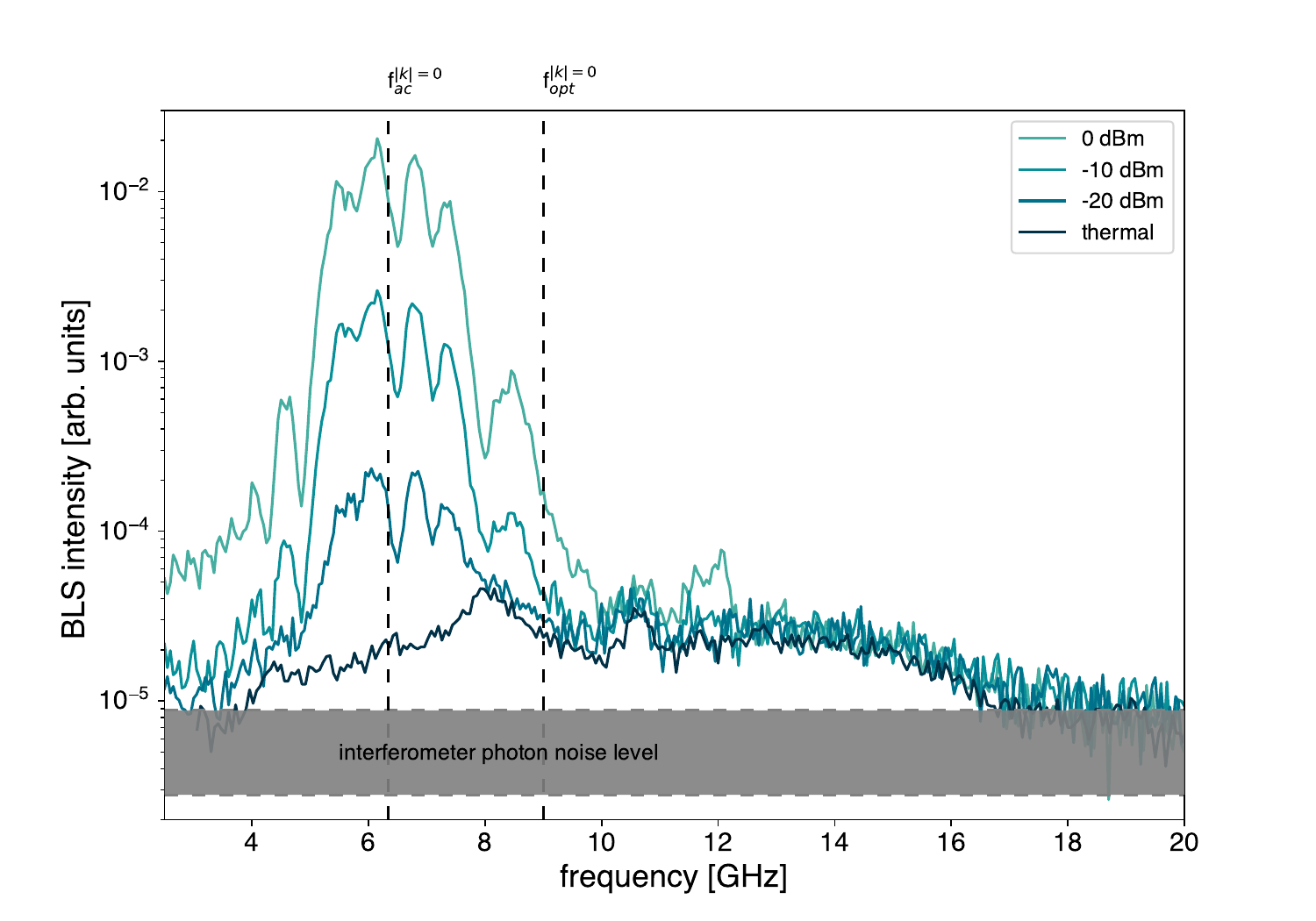}
\caption{Microfocused BLS spectra recorded at a fixed position at the middle of the SAF stripe for a field of 50 mT. Gray curve: Thermal spin waves. Green curves: Antenna-driven spin waves for applied powers of -20, -10 and 0 dBm. The vertical dotted lines are the $\vec k =\vec 0$ points of the unbounded SAF.}
\label{Figure_5_thermalBLS_vs_drivenBLS}
\end{figure}

\subsection{Supplementary: Non-excitation of the optical spin waves} 
The orientation of the r.f. field of the antenna obeys $\vec{h}^\textrm{rf} \perp \vec H_x$, such that the optical spin waves of an unbounded SAF film cannot be excited. Whether this non-excitation of the optical spin waves still applies for a stripe of finite width is answered thanks to Fig.~\ref{Figure_5_thermalBLS_vs_drivenBLS}. \\
This figure gathers microfocused BLS spectra: the spectra are taken with a laser focused at the diffraction limit at a \textit{fixed} position at the midst of the stripe, 3 $\mu$m away from the antenna edge. In this configuration, all SWs present in the sample contribute to the signal, whatever the direction of their wavevector but provided their modulus is smaller than $k_\textrm{max}^\textrm{BLS}=18 ~\textrm{rad}/\mu$m. 

The spectrum of thermal spin waves [dark blue curve in Fig.~\ref{Figure_5_thermalBLS_vs_drivenBLS}] is recorded without powering the antenna. There is BLS signal in frequency regions corresponding to both the acoustic and the optical spin waves.
Conversely, the other spectra in Fig.~\ref{Figure_5_thermalBLS_vs_drivenBLS} are recorded when powering the antenna with an r.f. source whose frequency tracks the measured BLS frequency. The power of the r.f. source (10, 100 $\mu$W and 1 mW) ensures that the driven SWs lead to signals that peak typically 1-3 orders of magnitude above that of the thermal SWs. The comparison of the spectra indicates that the additional spin waves driven by the antenna are purely of acoustic nature [Fig.~\ref{Figure_5_thermalBLS_vs_drivenBLS}(b)] except at very large powers. We can thus analyze the antenna-driven BLS images of Fig.~2 using essentially the sole contribution of the acoustic spin waves.
%
\begin{figure}
\includegraphics[width=\linewidth]{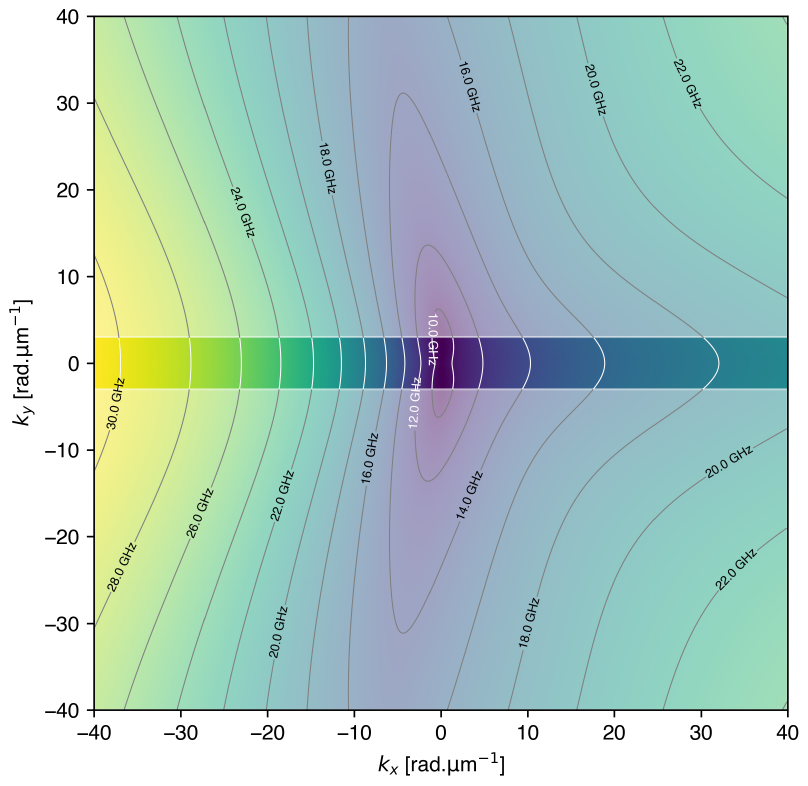}
\caption{Dispersion relations of the optical spin waves in an infinitely extended SAF film at a field of $\mu_0 H_x$=50~mT.}
\label{Figure_6_opticalDispersionRelations_v3}
\end{figure}

\subsection{Supplementary: Dispersion relation of the optical spin waves} 
The dispersion relation of the optical SWs is reported in Fig.~\ref{Figure_6_opticalDispersionRelations_v3} for a field of 50 mT in an unbounded film. The frequencies of these optical SWs are above any of the acoustic ones considered above, which is an additional argument to disregard the optical SWs when interpreting our experimental results. Their dispersion is also very different. The isofrequency lines run mostly parallel to the $k_y$ direction, meaning that the optical SWs have a group velocity with a component in the $y$ (guided) direction that almost vanishes.  In the other direction, the isofrequency lines are very closely spaced, which accounts for large group velocities, here typically 11 km/s, way above that of acoustic spin waves (typically 2 km/s). This direction of the group velocity is such that an antenna is not expected to radiate energy through optical spin waves along the stripe when the field is applied in such transverse direction. The optical spin waves have a non-reciprocal character inducing $\omega_\textrm{op}(k_x, k_y) \neq \omega_\textrm{op}(-k_x, k_y)$ but they are still forward waves: they have a standard energy-propagating character [i.e. $\frac{\partial\, \omega_\textrm{op}} {\partial\, |k_x|} \Big \rvert_{k_y=0} > 0$].

\subsection{Supplementary: Micromagnetic simulations of the BLS images for other stripe widths} 

The Figs.~7 and 8 report a selection of BLS images predicted by micromagnetics (see Eq.~2 in methods) for stripe widths ranging from 1 to 10 $\mu$m. 

%
\begin{figure*}
\includegraphics[width=\linewidth]{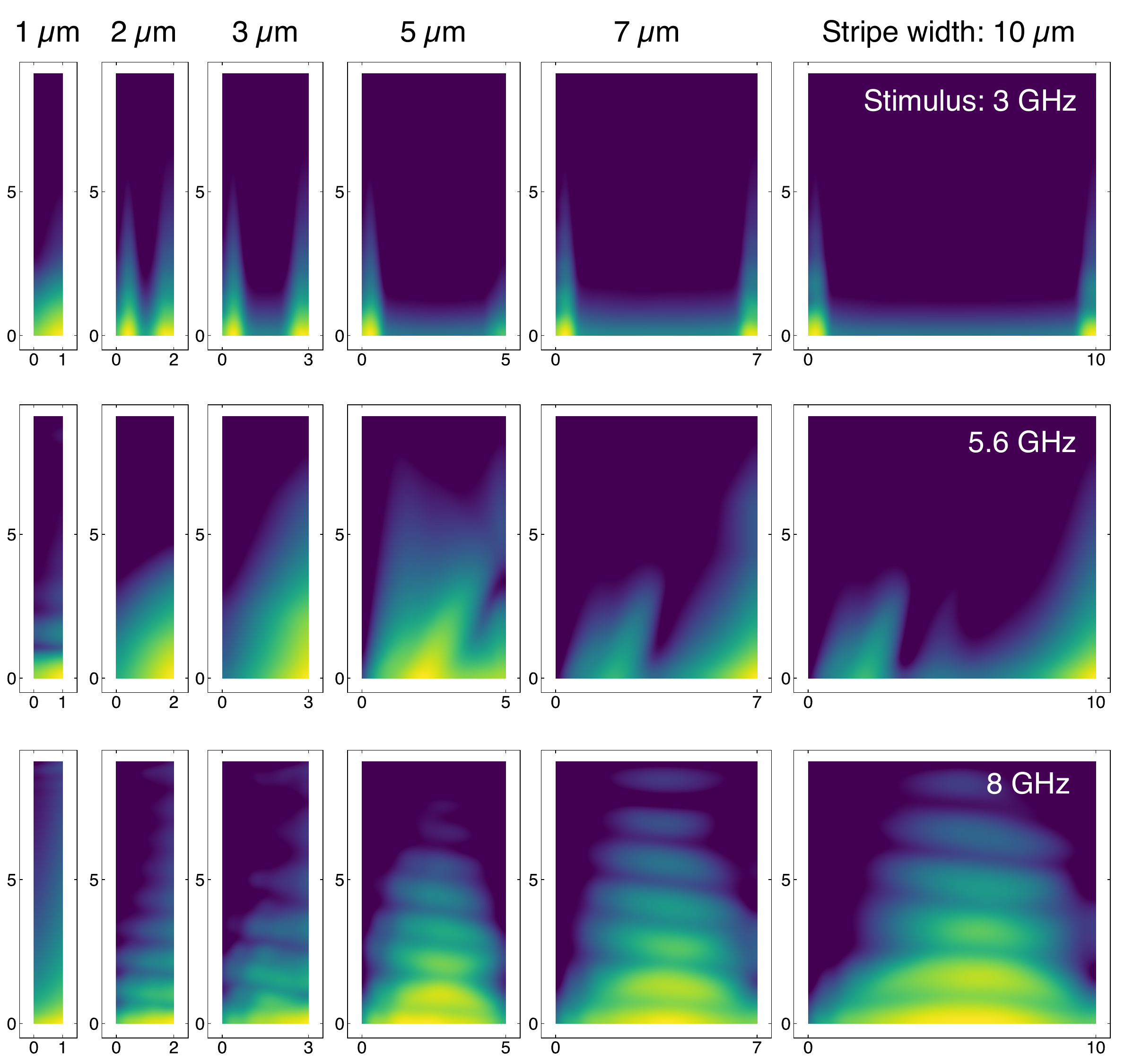}
\caption{Micromagnetic simulations of the BLS images after convolution with a Gaussian filter reducing the lateral resolution to 350 nm. The color scale is logarithmic. Left to right: for stripe widths being 1, 2, 3, 5, 7 and 10 $\mu$m. Top to Bottom: for stimulus frequencies being 3, 5.6 and 8 GHz. )}
\label{FigureSupplementaryMicromagneticData}
\end{figure*}

%
\begin{figure*}
\includegraphics[width=\linewidth]{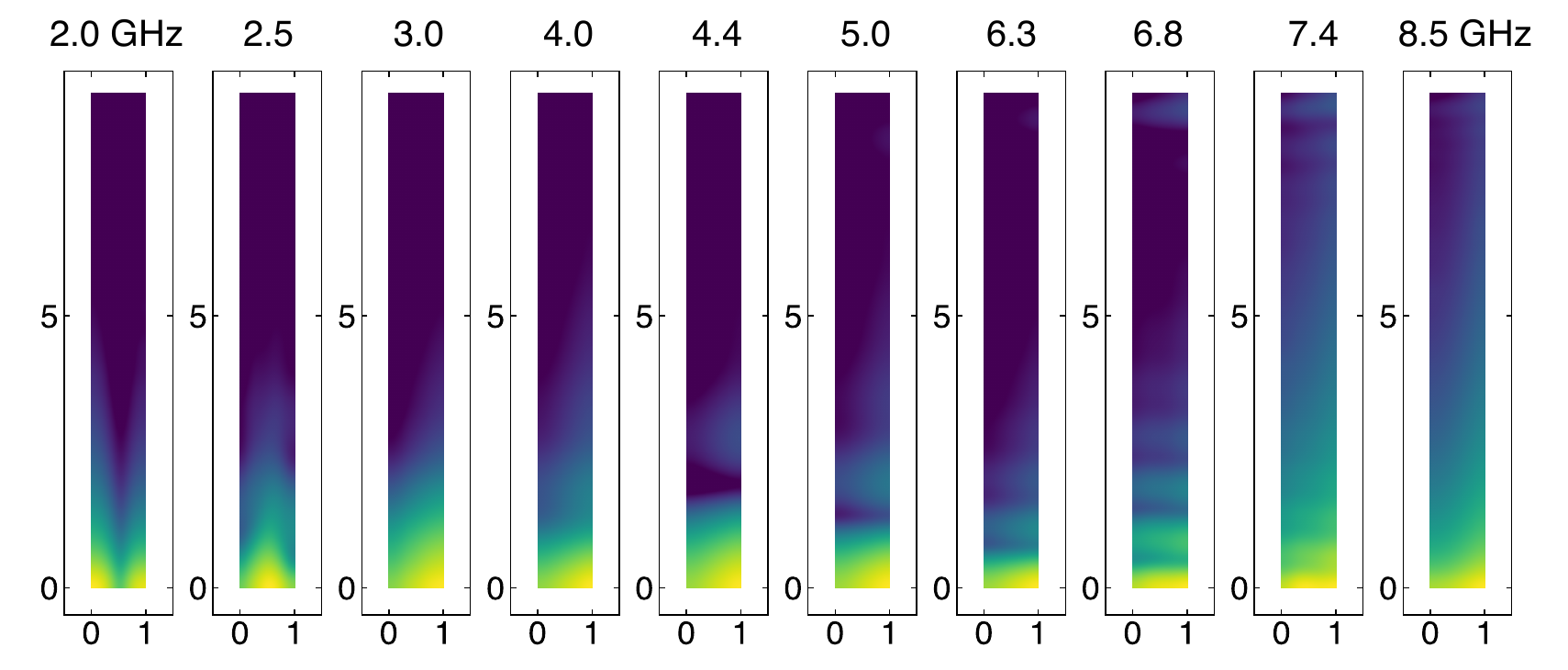}
\caption{Micromagnetic simulations of the BLS images for a stripe widths of 1 $\mu$m after convolution with a Gaussian filter reducing the lateral resolution to 350 nm. The color scale is logarithmic. Left to right: for increasing stimulus frequencies.)}
\label{FigureSupplementaryMicromagneticDataAmicronStripe}
\end{figure*}

\end{document}